\documentclass[epj,twocolumn,numbook]{svjour}
\usepackage{latexsym}
\usepackage{graphics}
\usepackage{dcolumn}
\usepackage{bm}
\usepackage{graphics}
\usepackage{graphicx}
\usepackage{epsfig}
\usepackage{booktabs,multirow}
\usepackage{hhline}
\usepackage{slashed}
\usepackage{rccol}
\usepackage{mathrsfs,amsmath,amssymb}
\usepackage{xcolor,colortbl}

\usepackage{xcolor}\colorlet{linkequation}{blue}
\usepackage[colorlinks=true, 
            linkcolor=red,   
            filecolor=red,   
            citecolor=blue, 
            urlcolor=blue,
            hyperfootnotes=true]{hyperref}

\newcommand\ba{\begin{eqnarray}}
\newcommand\ea{\end{eqnarray}}

\begin{document}
\title{Analytical bound state solutions of the Dirac equation with the Hulth\'en plus a class of Yukawa potential including a Coulomb-like tensor interaction}
\author{A.~I.~Ahmadov\inst{1,3}\thanks{ahmadovazar@yahoo.com}  \and M. Demirci\inst{2}\thanks{mehmetdemirci@ktu.edu.tr (Corresponding author)}\and M. F. Mustamin\inst{2} \and S. M.~Aslanova\inst{1} \and M.~Sh.~Orujova\inst{4}
%
}                     
%
\institute{Department of Theoretical
Physics, Baku State University, Z. Khalilov st. 23, AZ1148, Baku,
Azerbaijan \and Department of Physics,
Karadeniz Technical University, TR61080, Trabzon, Turkey \and Institute  for Physical Problems, Baku
State University, Z. Khalilov st. 23, AZ1148, Baku, Azerbaijan \and Azerbaijan State University
of Economics, Istiqlaliyyat st.6, AZ1001, Baku, Azerbaijan}
\date{Received: \today}
%
\abstract{We examine the bound state solutions of the Dirac equation under the spin and pseudospin symmetries for a new suggested combined potential, Hulten plus a class of Yukawa potential including a Coulomb-like tensor interaction. An improved scheme is employed to deal with the centrifugal (pseudo-centrifugal) term. Using the Nikiforov-Uvarov and SUSYQM methods, we analytically develop the relativistic energy eigenvalues and associated Dirac spinor components of wave functions. We find that both methods give entirely the same results. Modifiable of our results into some particular potential cases, useful for other physical systems, are also discussed. We obtain complete agreement with the findings of previous works. The spin and pseudospin bound state energy spectra for various levels are presented in the absence as well as the presence of tensor coupling. Both energy spectrums are sensitive with regards to the quantum numbers $\kappa$ and $n$, as well as the parameter $\delta$. We also notice that the degeneracies between Dirac spin and pseudospin doublet eigenstate partners are completely removed by the tensor interaction. Finally, we present the parameter space of allowable bound state regions of potential strength $V_0$ with constants for both considered symmetry limits $C_S$ and $C_{PS}$.
} 

\maketitle
\section{Introduction}
In relativistic quantum mechanics (QM), the Dirac equation is used to describe dynamics for many composite and non-composite subatomic systems that possess spin-$1/2$~\cite{Dirac,Herman,Bagrov1,Greiner}. The equation has been applied to investigate physical phenomena in a wide range of topics, especially in the nuclear and hadronic physics. In the advancement to these areas, two kinds of symmetries are introduced to the Dirac equation: spin and pseudospin \cite{Ginocchio0,Ginocchio1,Ginocchio2,Ginocchio3,Ring}. The spin symmetry produces two degeneracies of states with quantum numbers ($n, l, j = l\mp s$)\footnote{For clearance, $n, l, s$, and $j$ denote radial, orbital, spin and total angular momentum quantum numbers, respectively.}, allowing it to be considered as a spin-doublet. For the pseudospin symmetry case, there is a quasi-degeneracy from which the degenerate states have two units differences in orbital angular momentum: ($n,l,j=l+1/2$) and ($n-1, l+2, j=l+3/2$). These products can also be regarded as a pseudospin doublet with quantum numbers ($\tilde n=n-1,\tilde l=l+1,\tilde j=\tilde l\pm\tilde s$). Here, the pseudospin and pseudo-orbital angular momentum are denoted as $\tilde s= 1/2$  and  $\tilde l$, respectively \cite{Arima,Hecht}. Furthermore, the pseudo-orbital angular momentum can be interpreted as the orbital angular momentum of the Dirac spinor lower component \cite{Ginocchio0}.

Many works have been done on several applications of these two symmetries, such as explaining the antinucleon spectrum of a nucleus \cite{Ginocchio1,Ginocchio2,Ginocchio3,Ring}, the process of nuclear deformation \cite{Bohr} as well as nuclear superdeformation \cite{Dudek}, effective nucleus shell-model \cite{Trol}, and small spin-orbit splitting in hadrons \cite{Page}. Particularly, the pseudospin symmetry has been implemented with several kinds of potentials, like the harmonic oscillator \cite{Lisboa}, Woods-Saxon~\cite{Guo}, Hulth\'en \cite{Soylu,Ikhdair11,Ikhdair112,Haouat08}, Yukawa \cite{Aydogdu11,Ikhdair12,Pakdel}, Morse \cite{Berkdemir,Qiang},  Mie-type \cite{AydogduMie,HamzaviMie}, Pöschl-Teller \cite{Chen,Jia1} or the Manning-Rosen \cite{Gao,Yanar} and recently with hyperbolic-type potentials \cite{Karayer}. Moreover, the influence of tensor interaction potential on both types of symmetries shows that all doublets lose their degeneracies \cite{Tensor04}. In regards to this idea, the inclusion of tensor potentials for solving the Dirac equation has been carried out in many studies \cite{Akcay09,Aydogdu10,Hamzavi10,Ikot15,Mousavi}.

Accompanying scalar, vector and tensor interactions to explain various systems in an effective way to the Dirac equation may help us understand complication of nuclear structure. Investigating straightly with quantum chromodynamics at a full-scale seems to be hopelessly difficult at present. Many-body physics approach can be done, but still not offering a simple way. As a result, a simplified, mean-field type approach is always valuable. An extreme version of this would be to assume an effective potential generated by the full set of particles, and look at the energy levels of a single nucleon in this effective field. Since involved interactions might depend on the individual nucleon spin as well, the most general structure allows these scalar, vector and tensor potentials. Regarding these backgrounds, our main concern in the present study is to examine the bound state solutions of the Dirac equation under the spin and pseudospin symmetry limits with new suggested potential. The potential consists of the Hulth\'en~\cite{Hulten1,Hulten2}
\begin{equation}
V_H(r)=-\frac{Ze^2\delta e^{-\delta r}}{(1-e^{-\delta r})},
\label{Hulthen}
\end{equation}
plus a class of Yukawa-type potential
\ba
V_{CY}(r)=-\frac{A e^{-\delta r}}{r}-\frac{B e^{-2\delta r}}{r^2}. \label{Yukawa}
\ea
The parameter $Z$ is the atomic number, while $\delta$ and $r$ are the screening parameter and separation distance of the potential, respectively. The parameters $A$ and $B$ indicate the interaction strengths. The Hulth\'en potential is classified as short-range potentials, extensively applied to describe the continuum and bound states of the interaction systems. It has been implemented in atomic, nuclear, and particle physics, particularly to deal with strong coupling, so that significant role may emerge for a particle under this potential. The Yukawa potential~\cite{Yukawa}, on the other hand, is a well-known effective potential that has successfully described the strong interactions between nucleons. It is widely used in plasma physics, in which it represents the potential for a charged particle affected by weakly non-ideal plasma, and also in electrolytes and colloids. Briefly speaking, both potentials are two simple representations of the screened Coulomb potential, i.e., they include Coulombic behavior for small $r$ and decrease exponentially as $r$ increased.

Some previous attempts have provided satisfactory energy bound states by examining both potentials separately. However, no one has considered them as a linear combination so far. In this study, we propose their combination for the first time, in order to obtain bound state solutions of Dirac equation. Various mentioned phenomena earlier can be investigated by utilizing combination of them to give alternative perspectives. Explicitly, from the two mentioned potentials, we have
\begin{equation}
V(r)=-\frac{Ze^2\delta e^{-\delta r}}{(1-e^{-\delta r})}-\frac{A e^{-\delta r}}{r}-\frac{B e^{-2\delta r}}{r^2}.
\label{HulYuk}
\end{equation}
For the tensor interaction, we use the following Coulomb-like potential
\begin{equation}
U(r)=-\frac{H}{r},~H=\frac{Z_a Z_b}{4\pi\varepsilon_0},~r\geq R_c,
\label{eq:tensorpot}
\end{equation}
where $Z_a$ and $Z_b$ represent the charge of the projectile $a$ and the target nuclei $b$, respectively. $R_c$ denotes the Coulomb radius with value is $7.78$ fm. In this work, we discuss the relativistic bound states in the arbitrary $\kappa$-wave Dirac equation by using this new-proposed potential in order to provide a more subtle formulation of physical properties, particularly on the energy of bound and continuum states for any interacting quantum systems.

In examining this system, we use two different widely used methods. The first is the Nikiforov-Uvarov (NU) method \cite{Nikiforov} within ordinary QM. The procedure is based on solving a second-order linear differential equation by transforming it into a generalized hypergeometric-type form. The second is the Supersymmetry QM (SUSYQM) method \cite{Gendenshtein1,Gendenshtein2}. Supersymmetry itself emerged as an attempt to unify all basic interactions in nature, firstly identified to unify bosonic and fermionic sector. Various standard QM phenomena have been successfully formulated with this ambitious model \cite{Cooper1,Cooper2}. This method has also been implemented to obtain the spin and pseudospin solutions of the Dirac equation under various potentials (see, Refs.\cite{Zarrinkamar,Maghsoodi,Feizi}). In what follows, we provide the relativistic bound state solutions for the above mentioned-combined system, obtained by using both methods and compare their results. 

We begin our discussion by constructing the Dirac equation in Sec.~\ref{equ}. We separately present its spin symmetry case in Sec.~\ref{esp} and pseudospin symmetry case in Sec.~\ref{psp}. In Sec.~\ref{br}, we provide our analytic results. We examine bound state solutions for both symmetry cases by using the NU method in Sec.~\ref{HYNU} and then by using the SUSYQM method in Sec.~\ref{HYSUSY}. In Sec.~\ref{pc}, the reducibility of our results into some potential cases are discussed. After that, we provide the numerical predictions for the dependence of energy spectra on $\delta$, $n$, $\kappa$ as well as other potential parameters in Sec.~\ref{nr}. Lastly, we summarize our work and give concluding remarks in Sec.~\ref{cr}.

\section{Governing Equation}\label{equ}
In a relativistic description, the Dirac equation of a particle with mass $M$ influenced by a repulsive vector potential $V(\vec{r})$, an attractive scalar potential $S(\vec{r})$, and a tensor potential $U(r)$ can be expressed in the following general form (with units such that $\hbar = c = 1$)
\begin{equation}
\Big[ \vec{\alpha}\cdot \vec{p} + \beta\Big(M+S(\vec{r})\Big)-i\beta \vec{\alpha}\cdot \hat{r} U(r) \Big] \psi(r,\theta,\phi) = \Big[E-V(\vec{r})\Big] \psi(r,\theta,\phi),
\label{a4}
\end{equation}
where $\vec{p}=-i\vec{\nabla}$ and $E$ are respectively the momentum operator and the relativistic energy of the system. $\vec{\alpha}$ and $\beta$ are the $4\times 4$ Dirac matrices which are defined as
\begin{eqnarray}
\vec{\alpha}=\begin{pmatrix} 0 & \vec{\sigma} \\ \vec{\sigma} & 0 \end{pmatrix}, \qquad
\beta= \begin{pmatrix} I & 0 \\ 0 & -I\end{pmatrix},
\label{a5}
\end{eqnarray}
with $\vec{\sigma}$ and $I$ respectively are the $2\times 2$ Pauli spin matrices and $2\times 2$ unit matrix.

For a particle within a spherical field, we can define the spin-orbit coupling operator $\hat K = -\beta(\vec{\sigma}\cdot\vec{L} + 1)$ with eigenvalue $\kappa$ alongside the total angular momentum operator $\vec{J}$. This operator commutes with the Dirac Hamiltonian, regardless of the concerned symmetry cases. Moreover, it also constructs a complete set of conservative quantities with $H^2, K, J^2, J_z$. This quantum number is then used to label the eigenstates, rather than the (pseudo-)orbital angular momentum. In the case of spherical symmetry, the potentials $V(\vec{r})$  and $S(\vec{r})$ in Eq.\eqref{a4} are depend only on the radial coordinates, such that $V(\vec{r})=V(r)$  and $S(\vec{r})=S(r)$ where $r=|\vec{r}|$. The Dirac spinors in this regards can then be classified in accordance with $\kappa$ and $n$ as
\begin{eqnarray}
\psi_{n\kappa}(r,\theta,\phi)=\frac{1}{r}\left(\begin{array}{ll}F_{n\kappa}(r)Y_{jm}^{l}(\theta,\varphi)\\
iG_{n\kappa}(r)Y_{jm}^{\tilde l}(\theta,\varphi)\end{array} \right).
\label{a6}
\end{eqnarray}
In this equation, $F_{n\kappa}(r)$ represents the upper and $G_{n\kappa}(r)$ the lower components of the radial wave function. There is also the spin spherical harmonic function $Y_{jm}^{l}(\theta,\varphi)$ and its pseudospin counterpart as $Y_{jm}^{\tilde l}(\theta,\varphi)$. Here, $m$ denotes the angular momentum projection on the z-axis. For a given $\kappa=\pm1,\pm2,\ldots$, the spin and pseudospin cases have respectively $l=|\kappa+1/2|-1/2$ and $\tilde l =|\kappa-1/2|-1/2$ to indicate their orbital angular momentum, while $j=|\kappa|-1/2$ for their total angular momentum. To connect $\kappa$ with the other quantum numbers, we have for the spin symmetry
\begin{equation}
\kappa=\left\{
      \begin{array}{llcl}
        l=+(j+\frac{1}{2}), &~(p_{1/2},d_{3/2}, \text{etc.}),&j=l-\frac{1}{2},  & \hbox{unaligned spin}~(\kappa>0), \\
        -(l+1)=-(j+\frac{1}{2}), &~(s_{1/2},p_{3/2}, \text{etc.}),&j=l+\frac{1}{2},&  \hbox{aligned spin}~(\kappa<0),
      \end{array}
    \right.
\end{equation}
 and for the pseudospin symmetry
\begin{equation}
\kappa=\left\{
      \begin{array}{llcl}
        +(\tilde l +1)=(j+\frac{1}{2}), &~(d_{3/2},f_{5/2}, \text{etc.}),&j=\tilde l+\frac{1}{2},  & \hbox{unaligned spin}~(\kappa>0), \\
        -\tilde l=-(j+\frac{1}{2}), &~(s_{1/2},p_{3/2}, \text{etc.}),&j=\tilde l-\frac{1}{2},&  \hbox{aligned spin}~(\kappa<0).
      \end{array}
    \right.
\end{equation}

Using the following identities~\cite{Bjorken}
\begin{eqnarray}
\begin{split}
&\vec{\sigma}\cdot \vec{p}=\vec{\sigma}\cdot \hat{r} \left(\hat{r} \cdot \vec{p}+i \frac{\vec{\sigma}\cdot \vec{L}}{r} \right),\\
\label{sigmap}
&(\vec{\sigma}\cdot \vec{L}) Y_{jm}^{\tilde l}(\theta,\varphi)=(\kappa-1)Y_{jm}^{\tilde l}(\theta,\varphi),\\
&(\vec{\sigma}\cdot \vec{L}) Y_{jm}^{ l}(\theta,\varphi)=-(\kappa+1)Y_{jm}^{l}(\theta,\varphi),\\
&(\vec{\sigma}\cdot \hat{r}) Y_{jm}^{l}(\theta,\varphi)=-Y_{jm}^{\tilde l}(\theta,\varphi),\\
&(\vec{\sigma}\cdot \hat{r}) Y_{jm}^{\tilde l}(\theta,\varphi)=-Y_{jm}^{l}(\theta,\varphi)
\end{split}
\label{sigmaL}
\end{eqnarray}
inserting Eq.\eqref{a6} into Eq.\eqref{a4}, and then splitting the angular part for their two spinor components, we can write the radial coupled Dirac equations as
\begin{eqnarray}
\left(\frac{d}{dr}+\frac{\kappa}{r}-U(r)\right)F_{n\kappa}(r)=\Big(M+E_{n\kappa}- \Delta(r)\Big)G_{n\kappa}(r),
\label{a7}
\end{eqnarray}
and
\begin{eqnarray}
\left(\frac{d}{dr}-\frac{\kappa}{r}+U(r)\right)G_{n\kappa}(r)=\Big(M-E_{n\kappa}+\Sigma(r)\Big)F_{n\kappa}(r),
\label{a8}
\end{eqnarray}
where $\Delta(r) = V(r)-S(r)$ and $\Sigma = V(r) + S(r)$ have been used. Eliminating $F_{n\kappa}(r)$ and $G_{n\kappa}(r)$ between Eq.\eqref{a7} and Eq.\eqref{a8}, we obtain the following equations
\begin{eqnarray}
\begin{split}
	\bigg[\frac{d^2}{dr^2}-\frac{\kappa(\kappa+1)}{r^2}+\frac{2\kappa}{r}U(r)-\frac{dU(r)}{dr}-U^2(r)&-\big(M+E_{n\kappa}-\Delta(r)\big)\big(M-E_{n\kappa}+\Sigma(r)\big) \\
&+ \frac{\frac{d\Delta(r)}{dr}(\frac{d}{dr} + \frac{\kappa}{r}-U(r))}{M + E_{n\kappa} - \Delta(r)}\bigg] F_{n\kappa}(r)=0,
\end{split}
\label{a9}
\end{eqnarray}
\begin{eqnarray}
\begin{split}
\bigg[ \frac{d^2}{dr^2} - \frac{\kappa(\kappa - 1)}{r^2}+\frac{2\kappa}{r}U(r)+\frac{dU(r)}{dr}-U^2(r) &- \big(M + E_{n\kappa} - \Delta(r)\big)\big(M - E_{n\kappa} + \Sigma(r)\big) \\
&- \frac{\frac{d\Sigma(r)}{dr}(\frac{d}{dr} - \frac{\kappa}{r}+U(r))}{M - E_{n\kappa} + \Sigma(r)} \bigg] G_{n\kappa}(r) = 0,
\end{split}
\label{a10}
\end{eqnarray}
with $\kappa(\kappa+1)=l(l+1)$ and $\kappa(\kappa-1)=\tilde{l}(\tilde{l}+1)$. Two different limit cases can be specified for these two equations. The Eq.\eqref{a9} is known as the spin symmetry while the Eq.\eqref{a10} is the pseudospin symmetry case. In addition to the previously mentioned applications, these symmetries also play an important role in the magnetic moment and identical bands of nuclear structure.

\subsection{Spin Symmetry Limit}\label{esp}
The spin symmetry occurs as $d\Delta(r)/dr=0$, so that $\Delta(r)=C_S=\text{constant}$ \cite{Meng1,Meng2}, and hence Eq.\eqref{a9} becomes
\begin{equation}
	\begin{split}
	\bigg[\frac{d^2}{dr^2} - \frac{\kappa(\kappa+1)}{r^2} +\frac{2\kappa}{r}U(r)&-\frac{dU(r)}{dr}-U^2(r)- \Big(M+E_{n\kappa} - C_S\Big)\Sigma(r) \\
&+ \Big(E_{n\kappa}^{2} - M^{2} + C_S(M - E_{n\kappa}) \Big) \bigg]F_{n\kappa}(r)=0,
	\end{split}
	\label{a11}
\end{equation}
where $\kappa=l$ for $\kappa>0$ and $\kappa=-(l+1)$ for $\kappa<0$. The $E_{n\kappa}$ depends on $n$ and $l$, which is associated with the spin symmetry quantum number. From Eq.~\eqref{a7}, the lower-spinor component can be expressed as
\begin{equation}
	\begin{split}
	G_{n\kappa}(r)=\frac{1}{M+E_{n\kappa} - C_S} \Big(\frac{d}{dr}+\frac{\kappa}{r}-U(r)\Big)F_{n\kappa}(r),
	\end{split}
	\label{GnkSS}
\end{equation}
where, as $E_{n\kappa}\neq -M$ for $C_{S} = 0$ (exact spin symmetry), there exist only real positive energy spectrum.

\subsection{Pseudospin Symmetry Limit}\label{psp}
In this limit, $d\Sigma/dr=0$, so that $\Sigma(r)=C_{PS}=\text{constant}$ \cite{Meng1,Meng2}. The Eq.\eqref{a10} then becomes
\begin{equation}
\begin{split}
\bigg[\frac{d^2}{dr^2} - \frac{\kappa(\kappa-1)}{r^2} +\frac{2\kappa}{r}U(r)&+\frac{dU(r)}{dr}-U^2(r)+ \Big(M-E_{n\kappa} + C_{PS}\Big)\Delta(r) \\
&- \Big(M^{2} - E_{n\kappa}^{2} + C_{PS}(M +E_{n\kappa})\Big) \bigg] G_{n\kappa}(r)=0
\end{split}
\label{difPSS}
\end{equation}
where $\kappa=-\tilde l$ for $\kappa<0$, and $\kappa=\tilde l+1$ for $\kappa>0$. The $E_{n\kappa}$ depends on $n$ and $\tilde l$, associated with the pseudospin quantum numbers. Note that the case $\tilde l\neq 0$ produces degenerate states with $j = \tilde l \pm 1/2$. This is classified as $SU(2)$ pseudospin symmetry. From Eq.~\eqref{a8}, the corresponding upper-spinor component can be written as
\begin{equation}
	\begin{split}
	F_{n\kappa}(r)=\frac{1}{M-E_{n\kappa}+ C_{PS}} \Big(\frac{d}{dr}-\frac{\kappa}{r}+U(r)\Big)G_{n\kappa}(r)
	\end{split}
	\label{FnkPSS}
\end{equation}
where now, as $E_{n\kappa}\neq M $ for $C_{PS} = 0$ (exact pseudospin symmetry), there exist only real negative energy spectrum.

\section{Analytical Treatment: Bound State Solutions}\label{br}
In this section, we treat the Dirac equation under the influence of the proposed potential and find its bound state solutions through the NU and SUSYQM methods.

\subsection{Implementation of Nikiforov-Uvarov Method}\label{HYNU}
\subsubsection{Spin Symmetry Case}
We first consider the Eq.\eqref{a11}, which contains the Hulth\'en plus a class of Yukawa and also a Coulomb-like tensor potential. It can be solved exactly only for $\kappa=0$ and $\kappa=-1$ in the absence of tensor interaction ($H=0$), since the centrifugal term (proportional to $\kappa(\kappa+1)/r^2$) vanishes. In the case of arbitrary $\kappa$, an appropriate approximation needs to be employed on the centrifugal terms. We use the following improved approximation \cite{Greene} for $\delta r \ll 1$
\begin{eqnarray}
\frac {1}{r}\approx\frac{2\delta e^{-\delta r}}{(1-e^{-2\delta r})} \label{a12}, \qquad
\frac{1}{r^{2}} \approx \frac{4\delta^{2} e^{-2\delta r}}{(1-e^{-2\delta r})^2}.
\label{a13}
\end{eqnarray}
It provides good accuracy for a small value of potential parameters. This approximation scheme has been commonly used for tackling the same issue (see Refs.\cite{Ahmadov1,Ahmadov2}, references therein). Under the approximation\footnote{For convenience, we substitute $\delta \rightarrow 2\delta$ in the Hulth\'en potential.}, our combined potential becomes
\begin{equation}
V'(r) = - \frac{(V_{0}+V_{0}')e^{-2\delta r}} {1-e^{-2\delta r}}-\frac{B'e^{-4\delta r}} {(1-e^{-2\delta r})^2},
\label{appHY}
\end{equation}
with $V_{0}=2\delta Z e^2$, $V_{0}'=2A\delta$, and $B'=4B\delta^2$.
\begin{figure}[tbh]
    \begin{center}
\includegraphics[scale=0.42]{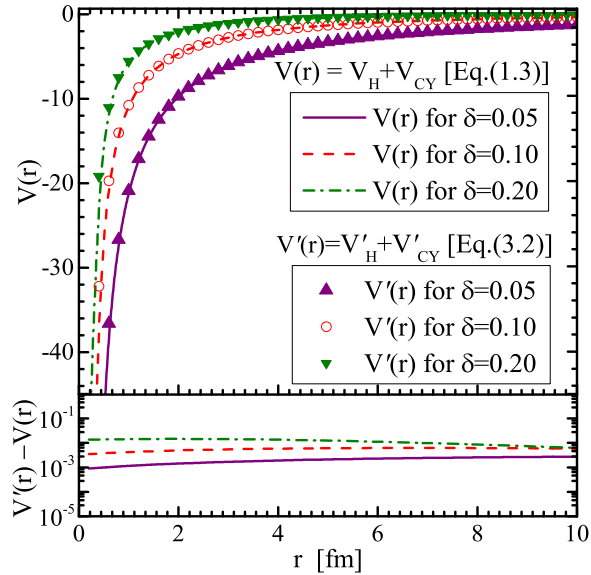}
     \end{center}
 \vspace{-4mm}
\caption{The effect of approximation on our potential as a function of separation distance $r$ for some values of parameter $\delta$.}
\label{fig:potapp}
\end{figure}
To quantitatively understand the approximation effect of the potential, the total potential~\eqref{HulYuk}, its approximation~\eqref{appHY} and difference between them as a function of $r$ for different values of $\delta$ are depicted in Fig.~\ref{fig:potapp}. Here, we set $V_0=2$ fm$^{-1}$, $A=1$ fm$^{-1}$ and $B=1$ fm$^{-1}$. It is obvious that for small $\delta$, the approximation becomes more suitable. The difference is about $10^{-3}$ and this is almost independent of $r$. It emphasizes that the equation~\eqref{a13} is a good approximation for centrifugal term as the parameter $\delta$ becomes small.

For the general form of the spin symmetry case,  we now consider the above approximation scheme and the tensor potential in Eq.~\eqref{eq:tensorpot}, so that Eq.\eqref{a11} becomes
\begin{equation}
\begin{split}
\bigg[\frac{d^2}{dr^2} - \frac{4\delta^{2}e^{-2\delta r}}{(1-e^{-2\delta r})^2}\Big(\kappa(\kappa+1) &+ 2\kappa H+H+H^2\Big)- \Big(M+E_{n\kappa}-C_S\Big) \Sigma(r)  \\
&+ \Big(E_{n\kappa}^{2} - M^{2} + C_S(M-E_{n\kappa})\Big) \bigg] F_{n\kappa}(r) = 0,
\end{split}
\label{a14}
\end{equation}
where $\Sigma(r)$ is taken as the potential~\eqref{appHY}.

Introducing $s=e^{-2\delta r}$ for $r \in[0,\infty)$ and $s\in[0,1]$, we can express the general form as
\begin{equation}
\begin{split}
\frac{d^2F_{n\kappa}}{ds^2} + \frac{1}{s}\frac{dF_{n\kappa}}{ds} + \bigg[& \frac{(V_0+V_{0}')} {4\delta^2s(1-s)} (M+E_{n\kappa}-C_S)+\frac{1}{4\delta^2s^2}\big(E_{n\kappa}^2 - M^2  + C_S(M - E_{n\kappa})\big)  \\
&+\frac{B' (M+E_{n\kappa}-C_S)} {4\delta^2(1-s)^2}- \frac{2\kappa H+H+H^2+\kappa(\kappa + 1)}{s(1-s)^2} \bigg] F_{n\kappa}=0.
\end{split}
\label{a16}
\end{equation}
We can further simplify this by defining
\begin{equation}
\begin{split}
&\alpha^2=\frac{(V_0+V_{0}')(M+E_{n\kappa}-C_S)}{4\delta^2}, \\
&\beta^2=\frac{M^2-E_{n\kappa}^2-C_S(M-E_{n\kappa})}{4\delta^2},\\
&\gamma^2=-\frac{B'(M+E_{n\kappa}-C_S)}{4\delta^2},\\
&\eta_\kappa=\kappa+H,
\end{split}
\label{a17}
\end{equation}
thus we arrive at the following form
\begin{equation}
\begin{split}
F_{n\kappa}''(s) + \frac{1-s}{s(1-s)} F_{n\kappa}'(s) + \biggl[ \frac{1}{s(1-s)} \biggr]^2 \biggl[ \alpha^{2} s (1-s) - \gamma^{2} s^2- \beta^{2} (1-s)^2  - \eta_\kappa(\eta_\kappa+1)s \biggr] F_{n\kappa}(s) = 0.
\end{split}
\label{a19}
\end{equation}
The tensor potential generates a new spin–orbit centrifugal
term $ \eta_\kappa( \eta_\kappa + 1)$. The solutions of this equation need to satisfy the boundary conditions, such as $F_{n\kappa}(0)=0$ at $s=1$ for $r\rightarrow 0$ and $F_{n\kappa}(\infty)\rightarrow 0$ at $s=0$ for $r\rightarrow \infty$.

The above equation can be easily solved by means of the NU method. At this stage, we follow the procedure presented in appendix~\ref{NUformalism}. Firstly, comparing Eq.\eqref{a19} with Eq.\eqref{NU}, we obtain
\begin{eqnarray}
\begin{split}
&\tilde{\tau}(s)= 1-s,\\
&\sigma(s)=s(1-s), \\
&\tilde{\sigma}(s)= \alpha^{2}s(1-s)-\beta^2 (1-s)^2-\gamma^2 s^2-\eta_\kappa(\eta_\kappa+1)s.
\end{split}
\label{a20}
\end{eqnarray}
Following factorization in \eqref{NUfact}, we then have
\begin{eqnarray}
F_{n\kappa}(s)=y_{n\kappa}(s)\phi(s),
\label{FnkFac}
\end{eqnarray}
so that Eq. \eqref{a19} can be reduced to a hypergeometric type equation like in Eq.\eqref{NUhyper}, and then $y_{n\kappa}(s)$ can be identified as one of its solutions. Considering the condition in Eq.\eqref{NUphi} for the suitable function $\phi(s)$, we obtain from relation \eqref{NUphii}
\begin{equation}
\pi(s)=\frac{{-s}}{2} \pm \sqrt{(a-k)s^{2}-(b-k)s+c},
\label{a27}
\end{equation}
with
\begin{eqnarray}
\begin{split}
 &a=\frac{1}{4}+\alpha^{2}+\beta^{2}+\gamma^{2}, \\
 &b=\alpha^{2}+2\beta^{2}-\eta_\kappa(\eta_\kappa+1), \\
 &c=\beta^{2}.
\end{split}
\end{eqnarray}
If the discriminant of Eq.\eqref{a27} inside the square root is zero, the constant $k$ can be classified as
\begin{equation}
k_{\pm}=(b-2c) \pm 2\sqrt{c(a-b)+c^{2}}.
\label{k12}
\end{equation}
From Eq.\eqref{a27}, we obtain the following possibilities for each value of $k$
\begin{equation}
\pi(s)=\frac{{-s}}{2} \pm
\left\{
\begin{array}{l}
\Big(\sqrt{c}-\sqrt {c+a-b}\Big)s-\sqrt{c} \quad \text{for}\quad k_+, \\
\Big(\sqrt{c}+\sqrt {c+a-b}\Big)s-\sqrt{c} \quad \text{for}\quad k_-. \\
\end{array} \right.
\label{a25}
\end{equation}
Here we found four possible values of $\pi(s)$ from the NU method. The one with negative derivation, the $k_-$ case, is redefined as $\tau(s)$ while the other can be neglected since lack of physical significance. We then obtain
\begin{equation}
\pi(s)=\sqrt{c}-s \left(\frac{1}{2}+\sqrt{c}+\sqrt{a-b+c}\right),
\label{a29}
\end{equation}
\begin{equation}
\tau(s)=1+2\sqrt{c}-2s
\left(1+\sqrt{c}+\sqrt{a-b+c}\right). \label{a30}
\end{equation}
Considering all of these and using Eq.\eqref{NUgeneigen}, we find the eigenvalue as
\begin{equation}
\begin{split}
\lambda=b-2c-2\sqrt{c^2+c(a-b)}-\left(\frac{1}{2}+{\sqrt{c}+\sqrt{a-b+c}}\right).
\label{a32}
\end{split}
\end{equation}
The hypergeometric-type equation provides a unique $n$-degree polynomial solution for non-negative integer $n$ like in Eq.\eqref{NUeigen}, with $\lambda_m\neq \lambda_n $ for $m=0,1,2,...,n-1$. Consequently
\begin{equation}
\begin{split}
\lambda_{n}&= 2n\left[{1 + \left({\sqrt{c} + \sqrt{a-b+c}} \right)}\right]+n(n-1).
\end{split}
\label{lambdaN}
\end{equation}
By inserting Eq.\eqref{a32} into Eq.\eqref{lambdaN} and explicitly solving this with $c=\beta^2$, we find
\begin{equation}
\begin{split}
M^2-&E_{n \kappa}^{2} - C_S(M - E_{n \kappa}) =\left[\frac{\alpha^{2}-\eta_\kappa(\eta_\kappa+1)-1/2-n(n+1)-(2n+1)\sqrt{\frac{1}{4}+\gamma^2+\eta_\kappa(\eta_\kappa+1)}}{n+\frac{1}{2}+\sqrt{\frac{1}{4}+\gamma^2+\eta_\kappa(\eta_\kappa+1)}}\delta\right]^2,
\label{EnSS}
\end{split}
\end{equation}
where $n=0,1,2,...$, $M>E_{n \kappa}$ and $E_{n \kappa}+M>C_S$. Note that the expression under the square root $\geq 0$. Otherwise, we have no bound state solutions. Here we have two different energy solutions for each value of $n$ and $\kappa$. However, the valid solution is the one that gives positive-energy bound states \cite{Ginocchio3}. Furthermore, it is possible to have degenerate states with various quantum numbers $n$ and $\kappa$ (or $l$) with same energy eigenvalues as $H=0$.

Now, we attempt to obtain the associated wave function for the proposed potential. By placing $\pi(s)$ and $\sigma(s)$ into Eq.\eqref{NUphi}, and then solving the first order differential equation, one part of the factorization is found to be
\begin{equation}
\phi (s)=s^{\beta}(1-s)^{\xi},
\label{phi0}
\end{equation}
with
\begin{equation}
\xi=\frac{1}{2}+\sqrt{\frac{1}{4}+\gamma^2+\eta_\kappa(\eta_\kappa+1)}.
\label{xi0}
\end{equation}
The other part with the hypergeometric-type function $y_{n\kappa}(s)$ has polynomial solutions that can be obtained from the Rodrigues relation of Eq.\eqref{NUrod}. Solving Eq.\eqref{NUweight} for the spin symmetry case, we find
\begin{equation}
\rho(s)=s^{2\beta}(1-s)^{2\xi-1},
\label{a42}
\end{equation}
and then substituting this into Eq.~\eqref{NUrod}, we obtain
\begin{equation}
\begin{split}
y_{n \kappa}(s)&=C_{n \kappa}(1-s)^{-2\xi+1}s^{-2\beta} \frac{{d^{n}}}{{ds^{n}}}\left[{s^{{2\beta}+n}
	(1-s)^{2\xi-1+n}}\right].
\end{split}
\label{a43}
\end{equation}
It is possible to simplify this by introducing the Jacobi polynomials \cite{Abramowitz}
\begin{equation}
\begin{split}
P_n^{(a,b)}(s)&=\frac{(-1)^n}{2^n n!(1-s)^a(1+s)^b}\frac{d^n}{ds^n
}\left[{(1-s)^{a+n}(1+s)^{b+n}}\right].
\label{a44}
\end{split}
\end{equation}
From this relation we find
\begin{equation}
P_n^{(a,b)}(1-2s)=\frac{1}{n!s^a(1-s)^b}\frac{d^{n}}{ds^{n}}\left[s^{a+n}(1-s)^{b+n}\right],
\label{a45}
\end{equation}
thus
\begin{equation}
\frac{d^n }{ds^n
}\left[s^{a+n}(1-s)^{b+n}\right]=n!s^{a}(1-s)^{b}
P_n^{(a,b)}(1-2s). \label{a46}
\end{equation}
By comparing the last expression with Eq.~\eqref{a43}, one gets
\begin{equation}
y_{n \kappa}(s) = C_{n \kappa} P_{n}^{(2\beta,2\xi-1)} (1-2s).
\label{ynk0}
\end{equation}
Putting $\phi(s)$ of Eq.\eqref{phi0} and $y_{n \kappa}(s)$ of Eq.\eqref{ynk0} into Eq.\eqref{FnkFac} leads to
\begin{equation}
F_{n \kappa}(s)=C_{n \kappa}s^{\beta}(1-s)^{\xi}
P_{n}^{(2\beta,2\xi-1)}(1-2s).
\label{a48}
\end{equation}
Implementing the identity of Jacobi polynomials \cite{Abramowitz}
\begin{equation}
\begin{split}
P_n^{(a,b)}&(1-2s)=\frac{{\Gamma(n+a+1)}}{{n!\Gamma(a+1)}} \mathop
{_{2}F_{1}} \left({-n,a+b+n+1,1+a;s}\right),
\label{a49}
\end{split}
\end{equation}
we can write down the upper component of the spinor in terms of the hypergeometric polynomial as
\begin{equation}
\begin{split}
F_{n \kappa}(s)=&C_{n \kappa}s^{\beta} (1-s)^{\xi}\frac{\Gamma (n + 2\beta+1)}{n! \Gamma (2\beta+1)} \mathop {_{2}F_{1}} \left({-n , {2\beta} + 2\xi + n, 1 + {2\beta} ; s} \right).
	\label{a50}
\end{split}
\end{equation}
One can obtain $F_{n \kappa}(r)$ by defining $s=e^{-2\delta r}$ in the above equation. We can implement this function, expressed with $r$-variable, into the Eq.\eqref{GnkSS} to obtain the corresponding lower component as
\begin{eqnarray}
\begin{split}
G_{n \kappa}(r)=\frac{D_{n \kappa}}{M+E_{n\kappa}-C_S}&\bigg[\frac{2 \delta n (2 \beta +2 \xi+n)\left(e^{-2 \delta  r}\right)^{\beta +1}}{(2 \beta +1)\left(1-e^{-2 \delta  r}\right)^{-\xi}}  \mathop {_{2}F_{1}}\left(1-n,2\xi+n+2\beta +1,2\beta +2;e^{-2\delta  r}\right)\\
&+\left(\frac{2 \delta  \xi  e^{-2 \delta r}}{1-e^{-2 \delta  r}}-2 \beta  \delta +\frac{\kappa+H}{r}\right) F_{n\kappa}(r)\bigg].
	\label{GnkSS2}
\end{split}
\end{eqnarray}
Finally, using the following normalization
\begin{equation}
\begin{split}
\int\limits_0^\infty
|R(r)|^{2}r^{2}dr&=\int\limits_0^{\infty} |\chi(r)|^{2}
dr=\frac{1}{2\delta}\int\limits_0^1\frac{1}{s}|\chi (s)|^{2}ds=1,
\label{Cnormal}
\end{split}
\end{equation}
and the following integral identity \cite{Abramowitz}
\begin{equation}
\begin{split}
\int\limits_0^1 &{dz(1-z)^{2(\nu+1)}z^{{2\mu}-1}}
\biggl[{\mathop {_{2}F_{1}}
	(-n,2(\nu+\mu+1)+n,2\mu+1;z)} \biggr]^2 \\
& = \frac{{(n+{\nu}+1)n!\Gamma (n+{2\nu}+2)\Gamma
		(2\mu)\Gamma({2\mu}+1)}}{{(n+{\nu}+{\mu}+1)\Gamma(n+{2\mu}+1)\Gamma
		(2({\nu}+{\mu}+1)+n)}},
\label{Cintegral}
\end{split}
\end{equation}
with $\nu>-3/2$ and $\mu>0$, we obtain the normalization constant of the spin symmetric wave function as
\begin{equation}
C_{n \kappa}=\sqrt{\frac{2\delta n! (n+\xi+\beta) \Gamma({2\beta} + 1) \Gamma(n + {2\beta} + 2\xi)}{(n + \xi)\Gamma (2\beta)\Gamma (n + 2\beta+1)\Gamma(n+2\xi)}}.
\label{a53}
\end{equation}

\subsubsection{Pseudospin Symmetry Case}
We now consider the Eq.\eqref{difPSS} and will follow similar steps with the spin symmetry case. Same as before, the equation can not be solved exactly for $\kappa\neq0$ or $\kappa\neq1$ without tensor interaction. Applying the same approximations (Eq.\eqref{a12}) to the centrifugal terms of \eqref{difPSS}, the form of general differential equation for the pseudospin symmetry becomes
\begin{equation}
\begin{split}
\biggl[\frac{d^2}{dr^2} - \frac{4\delta^{2}e^{-2\delta r}}{(1 - e^{-2\delta r})^2}\Big(\kappa(\kappa-1) &+ 2\kappa H-H+H^2\Big) + \Big(M-E_{n\kappa}+C_{PS}\Big) \Delta(r) \\
 &- \Big(M^{2} - E_{n\kappa}^{2} + C_{PS}(M + E_{n\kappa})\Big) \biggr]G_{n\kappa}(r)=0,
\end{split}
\label{a55}
\end{equation}
where we consider the potential in Eq.\eqref{appHY} for $\Delta(r)$ and a Coulomb-like potential in Eq.~\eqref{eq:tensorpot} for the tensor interaction. We can simplify Eq.\eqref{a55} by defining $s=e^{-2\delta r}$ to obtain
\begin{equation}
\begin{split}
\frac{d^2G_{n\kappa}}{ds^2} + \frac{1}{s}\frac{dG_{n\kappa}}{ds} +& \bigg[\frac{-(V_0 + V_{0}')}{4\delta^2s(1-s)} (M - E_{n\kappa} +C_{PS}) +\frac{1}{4\delta^2s^2} \bigg(E_{n\kappa}^2 - M^2 - C_{PS}(M + E_{n\kappa})\bigg) \\
&-\frac{B'(M - E_{n\kappa}+C_{PS})}{4\delta^2(1-s)^2} -\frac{2\kappa H-H+H^2+\kappa(\kappa - 1)}{s(1 - s)^2} \bigg] G_{n\kappa}=0
\end{split}
\label{a57}
\end{equation}
Using the following definitions
\begin{equation}
\begin{split}
&\tilde{\alpha}^2 = - \frac{(V_0 + V_{0}')(M - E_{n\kappa} + C_{PS})}{4\delta^2}, \\
&\tilde \beta^2 = \frac{M^2-E_{n\kappa}^2 + C_{PS}(M + E_{n\kappa})}{4\delta^2},\\
&\tilde{\gamma}^2 = \frac{B'(M - E_{n\kappa} + C_{PS})}{4\delta^2},\\
&\tilde{\eta}_\kappa=\kappa+H,
\label{betapseudo}
\end{split}
\end{equation}
we can rewrite Eq.\eqref{a57} as
\begin{equation}
\begin{split}
&\frac{d^2G_{n\kappa}}{ds^2}+\frac{1}{s}\frac{dG_{n\kappa}}{ds}+\frac{1}{s^2(1-s)^2}\bigg[\tilde\alpha^2
s(1-s)-\tilde\gamma^2s^2-\tilde\beta^2(1-s)^2-\tilde{\eta}_\kappa(\tilde{\eta}_\kappa-1)s\bigg]G_{n\kappa}=0.
\label{a60}
\end{split}
\end{equation}
As the previous treatment, the solution is restricted in the boundary conditions $G_{n\kappa}(0)=0$ at $s=1$ for $r\rightarrow 0$ and $G_{n\kappa}(\infty)\rightarrow 0$ at $s=0$ for $r\rightarrow \infty$. Comparing Eq.\eqref{a60} with Eq.\eqref{NU}, we obtain
\begin{eqnarray}
\begin{split}
&\tilde{\tau}(s)= 1-s,\\
&\sigma(s)=s(1-s), \\
&\tilde{\sigma}(s)= \tilde\alpha^{2}s(1-s) - \tilde\beta^2 (1-s)^2 -\tilde\gamma^2 s^2- \tilde{\eta}_\kappa(\tilde{\eta}_\kappa-1)s,
\end{split}
\label{a61}
\end{eqnarray}
which have similar form with Eq.\eqref{a20} and only differ in the last parameter.

We then factorize the general solution as
\begin{eqnarray}
G_{n\kappa}(s)= \tilde y_{n\kappa}(s) \tilde\phi(s),
\label{a62}
\end{eqnarray}
and by using Eq.\eqref{a61} as well as Eq.\eqref{NUphi}, we find
\begin{equation}
\pi(s)=\frac{{-s}}{2} \pm \sqrt{(\tilde{a}-k)s^{2}-(\tilde{b} - k)s+\tilde{c}},
\label{a66}
\end{equation}
where
\begin{equation}
\begin{split}
&\tilde{a}=\frac{1}{4}+ \tilde{\alpha}^{2} + \tilde{\beta}^{2}+\tilde\gamma^2,\\
&\tilde{b}=\tilde{\alpha}^{2} + 2\tilde\beta^{2} - \tilde{\eta}_\kappa(\tilde{\eta}_\kappa-1),\\
&\tilde{c}=\tilde\beta^{2}.
\end{split}
\end{equation}
Following the same procedures as in Eq.\eqref{k12} - Eq.\eqref{lambdaN}, we obtain
\begin{equation}
\begin{split}
M^2-E_{n \kappa}^{2} &+ C_{PS}(M + E_{n \kappa}) =\left[\frac{\tilde{\alpha}^{2}-\tilde{\eta}_\kappa(\tilde{\eta}_\kappa-1)-\frac{1}{2}-n(n+1)-(2n+1)\sqrt{\frac{1}{4}+\tilde{\gamma}^2+\tilde{\eta}_\kappa(\tilde{\eta}_\kappa-1)}}{n+\frac{1}{2}+\sqrt{\frac{1}{4}+\tilde{\gamma}^2+\tilde{\eta}_\kappa(\tilde{\eta}_\kappa-1)}}\delta\right]^2,
\label{EnkPS}
\end{split}
\end{equation}
where $n=0,1,2,...$, $M>-E_{n \kappa}$ and $E_{n \kappa}<C_{PS}+M$. This relation shows that the pseudospin limit produces a quadratic eigenvalues as in the previous case. The bound state solutions can only be achieved if the expression inside the square root $\geq$ 0. Here we have two different energy solutions for each value of $n$ and $\kappa$. On the other hand, in this considered symmetry limit, only the negative energy eigenvalues are valid and there are no bound state from the positive ones ($E\neq M+C$) \cite{Ginocchio3}. Furthermore, we encounter degenerate states for various quantum numbers $n$ and $\kappa$ (or $\tilde l$) with the same energy spectrum as $H=0$.

Let us now examine the radial part of the eigenfunctions. By using Eq.\eqref{NUweight}, the corresponding weight function can be written as
\begin{equation}
\tilde \rho(s)=s^{2\tilde\beta}(1-s)^{2\tilde\xi-1},
\label{a79}
\end{equation}
with
\begin{equation}
\tilde\xi=\frac{1}{2}+\sqrt{\frac{1}{4}+\tilde{\gamma}^2+\tilde{\eta}_\kappa(\tilde{\eta}_\kappa-1)},
\label{a79b}
\end{equation}
so that
\begin{equation}
\begin{split}
\tilde y_{n \kappa}(s)&=C_{n \kappa}(1 - s)^{1-2\tilde\xi} s^{-2\tilde\beta} \frac{{d^{n}}}{{ds^{n}}} \left( {s^{{2\tilde\beta} + n} (1-s)^{2\xi-1+n}} \right).
\label{a80}
\end{split}
\end{equation}
Similar to the spin symmetry case, applying the Jacobi  polynomials gives us
\begin{equation}
\tilde y_{n \kappa}(s) = \tilde C_{n \kappa} P_{n}^{(2\beta,2\tilde\xi-1)} (1-2s).
\label{a84}
\end{equation}
From $\tilde \phi (s)$ and $\tilde y_{n \kappa}(s)$, the lower component of the spinor wave function becomes
\begin{equation}
G_{n \kappa}(s) = \tilde C_{n \kappa} s^{\tilde\beta} (1-s)^{\tilde\xi} P_{n}^{(2 \tilde \beta,2\tilde\xi-1)}(1-2s).
\label{a85}
\end{equation}
Proceed further using Eq.\eqref{a49}, we can express this equation in terms of the hypergeometric polynomial as
\begin{equation}
\begin{split}
G_{n \kappa} (s) =& \tilde C_{n \kappa}s^{\tilde\beta}(1-s)^{\tilde\xi} \frac{\Gamma (n + 2\tilde\beta+1)}{n! \Gamma (2\tilde\beta+1)} \mathop {_{2}F_{1}} \left({-n, {2\tilde\beta} + 2\tilde\xi+n, 1+{2\tilde\beta};s}\right).
\label{a87}
\end{split}
\end{equation}
We can express this as $r$-dependent equation by defining $s=e^{-2\delta r}$, and then by inserting this now $r$ dependent function into Eq.\eqref{FnkPSS}, we obtain the other component as
\begin{eqnarray}
\begin{split}
F_{n \kappa}(r)=\frac{\tilde D_{n \kappa}}{M-E_{n\kappa}+C_{PS}}&\bigg[\frac{2 \delta n (2 \tilde\beta +2 \tilde{\xi}+n)\left(e^{-2 \delta  r}\right)^{\tilde\beta +1}}{(2 \tilde\beta +1)\left(1-e^{-2 \delta  r}\right)^{-\tilde{\xi}}}  \mathop {_{2}F_{1}}\left(1-n,2\tilde{\xi}+2\tilde\beta+n +1,2\tilde\beta +2;e^{-2\delta  r}\right)\\
&+\left(\frac{2 \delta \tilde{\xi} e^{-2 \delta r}}{1-e^{-2 \delta  r}}-2 \tilde\beta  \delta -\frac{\kappa+H}{r}\right) G_{n\kappa}(r)\bigg].
	\label{FnkPS2}
\end{split}
\end{eqnarray}
Implementing normalization condition \eqref{Cnormal} and the integral identity \eqref{Cintegral}, the normalization constant of the pseudospin symmetry becomes
\begin{equation}
\tilde C_{n \kappa}=\sqrt{\frac{2\delta n!(n +\tilde\xi+\tilde\beta)\Gamma(2\tilde\beta+1)\Gamma(n +2\tilde\beta+2\tilde\xi)}{(n +\tilde\xi)\Gamma(2\tilde\beta) \Gamma(n+2\tilde\beta+1)\Gamma(n+2\tilde\xi)}}.
\label{a90}
\end{equation}

As a final remark of the NU method, notice that the following replacements
\begin{equation}
\begin{split}
&\kappa(\kappa+1)\leftrightarrow \kappa(\kappa-1)~(\text{or}~\kappa \leftrightarrow \kappa\pm1),\\
&F_{n \kappa}\leftrightarrow G_{n \kappa},~E^+_{n \kappa}\leftrightarrow -E^-_{n \kappa},\\
&(V_0+V_{0}')\leftrightarrow -(V_0+V_{0}'),C_S\leftrightarrow -C_{PS},
\\
&\beta^2 \leftrightarrow \tilde\beta^2,~\gamma^2 \leftrightarrow -\tilde\gamma^2,~\alpha^2 \leftrightarrow -\tilde\alpha^2,
\label{parameterexchaning}
\end{split}
\end{equation}
enable us to straightforwardly produced the negative energy solution of the pseudospin symmetry from the positive energy solution of the spin symmetry case. That is, Eqs.\eqref{EnSS} and \eqref{a50} give respectively Eqs.\eqref{EnkPS} and \eqref{a87} under the above replacements, or vice versa.

\subsection{Implementation of the SUSYQM Method}\label{HYSUSY}
Now, we are going to implement the SUSYQM method for both symmetry cases. The discussion follows conventions from appendix A of Ref.\cite{Ahmadov1}.

\subsubsection{Spin Symmetry Case}
According to the SUSYQM, the ground state of $F_{0}(r)$ in Eq.\eqref{a11} satisfies
\begin{equation}
F_{0}(r)=N\exp\left(-\int W(r)dr\right),
\label{a91}
\end{equation}
with normalization constant $N$. $W(r)$ is known as the superpotential and can be used to define the supersymmetric partner potentials \cite{Cooper1,Cooper2}
\begin{equation}
V_{\pm}(r)=W^{2}(r)\pm W'(r),
\label{SSpartner}
\end{equation}
which is also known as the Riccati equation. Here we take its particular solution as
\begin{equation}
W(r)=A-\frac{Be^{-2\delta r}}{1-e^{-2\delta r}},
\label{a93}
\end{equation}
with unknown constants $A$ and $B$. To find the solution of Eq.\eqref{a14} via SUSYQM, we rewrite the equation in general form as
\begin{equation}
\frac{d^2 F_{n \kappa}}{dr^2}= \Big(V_{\rm
eff}(r)-E\Big)F_{n \kappa}.
\label{susydifSS}
\end{equation}
Substituting $V_{-}(r)=V_{\rm eff}(r)-E_{0}$ ($E_{0}$ represents the ground-state energy)  and \eqref{a93} into Eq.\eqref{SSpartner}, and then comparing the compatible terms of the left- and right-hand sides, we obtain
\begin{eqnarray}
A^{2}&&=4\delta^{2}\beta^{2}, \label{a94} \\
2AB+2\delta B&&=4\delta^{2}\alpha^{2}-4\delta^{2}\eta_\kappa(\eta_\kappa+1),
\label{a95} \\
2 A B+B^2&&=4 \delta^2\left(\alpha^2+\gamma^2\right).\label{a96}
\end{eqnarray}
The requirements $B>0$ and $A<0$ are needed to describe the wave functions in extreme condition. Then, from Eqs.\eqref{a95} and \eqref{a96} we find the following relations
\begin{equation}
\begin{split}
B&=\frac{2\delta\pm\sqrt{4\delta^{2}+16\delta^{2}(\gamma^2+\eta_\kappa(\eta_\kappa+1))}}{2}=\delta\pm
2\delta\sqrt{\frac{1}{4}+\eta_\kappa(\eta_\kappa+1)+\gamma^2}.
\label{a97}
\end{split}
\end{equation}
\begin{equation}
A=-\frac{B}{2}+\frac{2\delta^{2}(\alpha^{2}+\gamma^2)}{B}.
\label{a98}
\end{equation}

We now back to Eq.\eqref{a93}. We approximate $W(r)\rightarrow$ $A$ as $r\rightarrow\infty$. Inserting Eq.\eqref{a93} into Eq.\eqref{SSpartner}, the supersymmetric partner potentials have the following forms
\begin{eqnarray}
\begin{split}
V_{-}(r)&=\biggl[A^2-\frac{(2AB+2\delta B)e^{-2\delta r}}{1-e^{-2\delta r}}+\frac{(B^2-2\delta B)e^{-4\delta r}}{(1-e^{-2\delta r})^2}\biggr], \label{a102} \\
V_{+}(r)&=\left[A^{2}-\frac{(2AB-2\delta	B)e^{-2\delta r}}{1-e^{-2\delta r}}+\frac{(B^2+2\delta B)e^{-4\delta r}}{(1-e^{-2\delta r})^2}\right]. \label{a103}
\end{split}
\end{eqnarray}
From these two equations, which is only differ from each other merely by additive constants, we can introduce their invariant forms as \cite{Gendenshtein1,Gendenshtein2}
\begin{eqnarray}
\begin{split}
R(B_1)&= V_{+}(B,r)-V_{-}(B_1,r)=\left[A^{2}-A_{1}^{2}\right] \\
&=\left[-\frac{B}{2}+\frac{2\delta^{2}(\alpha^{2}+\gamma^2)}{B}\right]^{2}-\left[-\frac{B+2\delta}{2}+\frac{2\delta^{2}(\alpha^{2}+\gamma^2)}{B+2\delta}\right]^2,
\label{a105}
\end{split}
\end{eqnarray}
or more generally
\begin{eqnarray}
\begin{split}
R(B_{i})&=V_{+}(B+(i-1)2\delta ,r)-V_{-}(B+i2\delta ,r) \\&= \Bigg[-\frac{B+(i-1) 2\delta}{2} + \frac{ 2 \delta^2 (\alpha^{2}+\gamma^2) } { B+(i-1) 2 \delta }\Bigg]^{2} - \Bigg[ -\frac{B+i 2\delta}{2}+\frac{2\delta^2 (\alpha^{2} + \gamma^2)} {B+i 2 \delta}\Bigg]^2.
\label{a106}
\end{split}
\end{eqnarray}
Continuing this procedure and substituting $\,B_{n}=B_{n-1} +2\delta =B+2n\delta$, the whole discrete spectrum of Hamiltonian $H_{-}(B)$ in general becomes
\begin{equation}
4\delta^2 \beta^2 =E_{0}^2+\sum_{i=1}^{n}R(B_i).
\label{a107}
\end{equation}
By setting $E_0=0$, we find
\begin{eqnarray}
\begin{split}
	4&\delta^2 \beta^2=\sum\limits_{i=1}^n R(B_i)\\
	&=\left(-\frac{B}{2} + \frac{2\delta^{2}(\alpha^{2} + \gamma^2)}{B}\right)^{2} - \left(-\frac{B}{2} + \frac{2 \delta^{2} (\alpha^{2} + \gamma^2)}{B}\right)^{2} \\
	& + \left( -\frac{ B+2\delta }{2} + \frac{ 2\delta^{2} (\alpha^{2}+\gamma^2)}{ B+2\delta }\right)^{2} - \left( -\frac{ B+2\delta }{2} + \frac{ 2\delta^{2}(\alpha^{2}+\gamma^2) }{ B + 2\delta }\right)^{2} \\
	& + \cdots - \\
	&+ \biggl( -\frac{B+2(n-1)\delta}{2}  + \frac{2\delta^{2} (\alpha^{2}+\gamma^2)}{ B + 2 (n-1)\delta}\biggr)^{2}- \left( -\frac{ B + 2 (n-1)\delta  }{2} + \frac{ 2 \delta^{2} (\alpha^{2}+\gamma^2)}{B+2(n-1)\delta} \right)^2 \\
	& + \left(-\frac{B + 2 n\delta}{2} + \frac{2\delta^{2}(\alpha^{2}+\gamma^2)}{(B+2 n\delta)} \right)^{2},
\end{split}
\label{a108}
\end{eqnarray}
so that
\begin{equation}
\beta^2=\frac{1}{4\delta^2}\left[-\frac{B+2 n \delta}{2} + \frac{2\delta^{2}(\alpha^{2}+\gamma^2)}{B+2n\delta} \right]^2.
\label{a109}
\end{equation}
By using Eq.\eqref{a97} and inserting $\beta^2$ of Eq.\eqref{a17} into this expression, we obtain the corresponding energy equation as
\begin{equation}
\begin{split}
M^2 -E_{n \kappa}^{2}- &C_S(M-E_{n \kappa})
\\
&=\delta^2\bigg[\frac{2\left(\alpha ^2+\gamma ^2\right)}{1+2n+2 \sqrt{\gamma^2+\eta_\kappa(\eta_\kappa+1)+\frac{1}{4}}}-\frac{1}{2}\left(1+2n+2 \sqrt{\gamma^2+\eta_\kappa (\eta_\kappa+1)+\frac{1}{4}}\right)\bigg]^2,
\label{a110}
\end{split}
\end{equation}
which is identical with the previous result of the NU method in Eq.\eqref{EnSS}.

From Eq.\eqref{a93}, the corresponding component of wave function can be written as
\begin{equation}
\begin{split}
	F_{0}(r)& = N \exp\left(-\int W(r)dr\right) \\
&= N \exp\left[\int\left(-A+\frac{Be^{-2\delta r}}{1-e^{-2\delta r}}\right)dr\right] \\
	& = Ne^{-Ar} \exp\left[\frac{B}{2\delta} \int\frac{d(1-e^{-2\delta r})}{1-e^{-2\delta r}}\right] \\
&= N e^{-Ar}(1-e^{-2\delta r})^{\frac{B}{2\delta}}.
	\label{a104}
\end{split}
\end{equation}
We can see that for $r\rightarrow 0$, $F_{0}(r)\rightarrow 0$ and $B>0$, while for $r\rightarrow \infty$, $F_{0}(r)\rightarrow 0$ and $A<0$.

\subsubsection{Pseudospin Symmetry Case}
The ground state of $G_{0}(r)$ in Eq.\eqref{a55} within SUSYQM can be written as
\begin{equation}
G_{0}(r)=\tilde N \exp\left(-\int\tilde W(r)dr\right),
\label{a111}
\end{equation}
where the normalization constant is now $\tilde N$. The supersymmetric partner potentials for the current consideration can be written as
\begin{eqnarray}
\tilde V_{\pm}(r)=\tilde W^{2}(r)\pm\tilde W'(r).
\label{a112}
\end{eqnarray}
The particular solution is now
\begin{equation}
\tilde W(r)= \tilde A-\frac{\tilde Be^{-2\delta r}}{1-e^{-2\delta r}},
\label{a113}
\end{equation}
with the unknown constants $\tilde A$ and $\tilde B$. We can rewrite Eq.\eqref{a55} in the following general form
\begin{equation}
\frac{d^2 G_{n \kappa}}{dr^2}= \Big(\tilde V_{\rm eff}(r)-\tilde{E}\Big)G_{n \kappa}.
\label{susydifPS}
\end{equation}
Substituting $\tilde V_{-}(r) = \tilde V_{\rm eff}(r)-\tilde{E}_{0}$ ($\tilde{E}_{0}$ represents the ground-state energy) and Eq.\eqref{a113} into Eq.\eqref{a112} we find
\begin{eqnarray}
\tilde A^{2}=&&4\delta^{2}\tilde\beta^{2}, \label{a114} \\
 2\tilde{A} \tilde{B} +2\delta \tilde
B=&&4\delta^{2}\tilde{\alpha}^{2}-4\delta^{2}\tilde{\eta}_\kappa(\tilde{\eta}_\kappa-1), \label{a115} \\
2 \tilde{A} \tilde{B}+\tilde{B}^2=&&4 \delta ^2\left(\tilde{\alpha}^2+\tilde{\gamma}^2\right). \label{a116}
\end{eqnarray}
The same argument with the spin symmetry leads to the condition $\tilde A<0$ and $\tilde B>0$, so that from Eq.\eqref{a115} and Eq. \eqref{a116} we obtain
\begin{equation}
\begin{split}
\tilde B=\frac{2\delta\pm\sqrt{4\delta^{2}+16\delta^{2}(\tilde{\gamma}^{2}+\tilde{\eta}_\kappa(\tilde{\eta}_\kappa-1))}}{2}
=\delta\pm
2\delta\sqrt{\frac{1}{4}+\tilde{\eta}_\kappa(\tilde{\eta}_\kappa-1)+\tilde{\gamma}^{2}},
\label{a117}
\end{split}
\end{equation}
\begin{equation}
\tilde A = -\frac{\tilde B}{2} + \frac{2\delta^{2} (\tilde\alpha^{2}+\tilde{\gamma}^{2})}{B}.
\label{a118}
\end{equation}

Back to Eq.\eqref{a113}, we approximate $\tilde W(r)\rightarrow$-$\tilde A$ as $r\rightarrow\infty$. Substituting Eq.\eqref{a113} into Eq.\eqref{a112} leads to
\begin{equation}
\begin{split}
\tilde{V}_{-}(r)&=\biggl[\tilde A^2-\frac{(2\tilde A\tilde B+2\delta \tilde B)e^{-2\delta r}}{1-e^{-2\delta r}}+\frac{(\tilde B^2-2\delta \tilde B)e^{-4\delta r}}{(1-e^{-2\delta r})^2}\biggr],\\
\tilde{V}_{+}(r)&=\left[\tilde A^{2}-\frac{(2\tilde A\tilde B-2\delta \tilde B)e^{-2\delta r}}{1-e^{-2\delta r}}+\frac{(\tilde B^2+2\delta\tilde B)e^{-4\delta r}}{(1-e^{-2\delta r})^2}\right].
\end{split}\label{a121}
\end{equation}
By using these relations, their invariant forms can be introduced as \cite{Gendenshtein1,Gendenshtein2}
\begin{equation}
\begin{split}
R (\tilde B_1) &= \tilde{V}_{+}(\tilde B,r) - \tilde{V}_{-}(\tilde B_1,r)=\left(\tilde A^{2}-\tilde A_{1}^{2}\right) \\
&=\bigg[\frac{\tilde B}{2}-\frac{2\delta^{2} (\tilde\alpha^{2}+\tilde{\gamma}^{2})}{\tilde B}\bigg]^{2} - \bigg[\frac{\tilde B+2\delta}{2}-\frac{2\delta^{2} (\tilde\alpha^{2}+\tilde{\gamma}^{2})}{\tilde B+2\delta}\bigg]^2,
\end{split}
\label{a124}
\end{equation}
or
\begin{eqnarray}
\begin{split}
R(\tilde B_{i})&= \tilde{V}_{+}\big(\tilde B+(i-1)2\delta ,r\big) - \tilde{V}_{-}\big(\tilde B+i2\delta ,r\big) \\
&= \bigg[\frac{\tilde B+(i-1)2\delta}{2}  -\frac{2(\tilde\alpha^{2}+\tilde{\gamma}^{2})\delta^2}{\tilde B + (i-1) 2\delta}\bigg]^{2} - \bigg[\frac{\tilde B+i2\delta}{2} - \frac{2(\tilde\alpha^{2}+\tilde{\gamma}^{2}) \delta^{2}}{\tilde B + i2\delta}\bigg]^2.
\end{split}
 \label{a125}
\end{eqnarray}
Continuing this and using $\tilde B_{n} = \tilde B_{n-1} + 2 \delta = \tilde B+2n\delta$, the complete spectrum of $H_{-}(\tilde B)$ becomes
\begin{equation}
4\delta^2\tilde \beta^2 = \tilde{E}_{0}^2 + \sum_{i=1}^{n}R(\tilde B_i).
\label{a126}
\end{equation}
As $\tilde{E}_0=0$, we obtain
\begin{eqnarray}
\begin{split}
4\delta^2 \tilde \beta^2&=\sum\limits_{i=1}^n R(\tilde B_i)\\
&=
\left(\frac{\tilde B}{2}-\frac{2\delta^{2} (\tilde\alpha^{2}+\tilde{\gamma}^{2})}{\tilde B}\right)^{2} - \bigg(\frac{\tilde B}{2} -\frac{2\delta^{2} (\tilde\alpha^{2}+\tilde{\gamma}^{2})}{\tilde B}\bigg)^{2} \\
&+\left(\frac{\tilde B+2\delta}{2} - \frac{2\delta^{2}(\tilde\alpha^{2}+\tilde{\gamma}^{2})}{\tilde B+2\delta}\right)^{2}- \left(\frac{\tilde B+2\delta}{2}-\frac{2\delta^{2} (\tilde\alpha^{2}+\tilde{\gamma}^{2})}{\tilde B+2\delta}\right)^{2} \\
&+ \cdots - \\
&+\biggl(\frac{\tilde B+2(n-1)\delta}{2} - \frac{2\delta^{2} (\tilde\alpha^{2}+\tilde{\gamma}^{2})}{\tilde B+2(n -1)\delta}\biggr)^{2} -\left(\frac{\tilde B + 2(n -1) \delta}{2}-\frac{2\delta^{2} (\tilde\alpha^{2}+\tilde{\gamma}^{2})}{\tilde B+2(n-1)\delta}\right)^2 \\
& +\left(\frac{\tilde B+2 n \delta}{2}-\frac{2\delta^{2} (\tilde\alpha^{2}+\tilde{\gamma}^{2})}{(\tilde B+2n \delta)}\right)^{2},
\end{split}
\label{a127}
\end{eqnarray}
and hence
\begin{equation}
\tilde\beta^2=\frac{1}{4\delta^2}\left( - \frac{\tilde B+2 n\delta}{2} + \frac{2\delta^{2} (\tilde\alpha^{2}+\tilde{\gamma}^{2})}{\tilde B+2 n\delta} \right)^2.
\label{a128}
\end{equation}
Using $\tilde\beta$ in Eq.\eqref{betapseudo} and $\tilde B$ of Eq.\eqref{a117}, the  energy spectrum equation becomes
\begin{equation}
\begin{split}
M^2-E_{n \kappa}^{2} &+ C_{PS}(M + E_{n \kappa})\\
&=\delta^2\bigg[\frac{2\left(\tilde{\alpha}^2+\tilde{\gamma}^2\right)}{1+2n+2\sqrt{\tilde{\gamma}^2+\tilde{\eta}_\kappa(\tilde{\eta}_\kappa-1) +\frac{1}{4}}}-\frac{1}{2}\left(1+2n+2\sqrt{\tilde{\gamma}^2+\tilde{\eta}_\kappa(\tilde{\eta}_\kappa-1)+\frac{1}{4}}\right)\bigg]^2,
\label{a129}
\end{split}
\end{equation}
which is identical to Eq.\eqref{EnkPS}. From the superpotential $\tilde{W}(r)$, we can express the eigenfunction $G_{0}(r)$ as
\begin{equation}
\begin{split}
G_{0}(r)&=\tilde{N} \exp\left(-\int \tilde{W}(r)dr\right) \\
&= \tilde{N} \exp \left[- \int\left(\tilde A - \frac{\tilde Be^{-2\delta r}}{1-e^{-2\delta r}}\right)dr\right] \\
&= \tilde{N} e^{-\tilde Ar} \exp\left[\frac{\tilde B}{2\delta}\int\frac{d(1-e^{-2\delta r})}{1-e^{-2\delta r}}\right]\\
&= N e^{-\tilde Ar}(1-e^{-2\delta r})^{\frac{\tilde B}{2\delta}},
\end{split}
\label{a123}
\end{equation}
where now, for $r\rightarrow 0$, $G_{0}(r)\rightarrow 0$ and $\tilde B>0$, whilst for $r\rightarrow \infty$, $G_{0}(r)\rightarrow 0$ and $\tilde A<0$.

\section{Particular cases}\label{pc}
Now, we are about to examine some particular cases regarding the bound state energy eigenvalues in Eq.\eqref{EnSS} and Eq.\eqref{EnkPS}. We could derive some well-known potentials, useful for other physical systems, by adjusting relevant parameters in both cases. We then compare the corresponding energy spectrums with the previous works.

\begin{description}
  \item[\textit{1. S-wave case}:] The s-wave cases are directly obtained for $l=0$ and $\tilde l=0$ ($\kappa=1$ for pseudospin symmetry and $\kappa=-1$ for spin symmetry), so that the spin–orbit coupling term vanishes. The corresponding energy eigenvalue equation reduces to
\ba
\begin{split}
M^2-E_{n, -1}^{2} - C_S(M - E_{n,-1})
=\delta^2\left[\frac{\alpha^{2}-1/2-H(H-1)-n(n+1)-(2n+1)\sqrt{\frac{1}{4}+\gamma^2+H(H-1)}}{n+\frac{1}{2}+\sqrt{\frac{1}{4}+\gamma^2+H(H-1)}}\right]^2
\end{split}
\ea
for the spin symmetry, and
\ba
\begin{split}
&M^2-E_{n,1}^{2} + C_{PS}(M + E_{n,1})
=\delta^2\left[\frac{\tilde{\alpha}^{2}-1/2-H(H+1)-n(n+1)-(2n+1)\sqrt{\frac{1}{4}+\tilde{\gamma}^2+H(H+1)}}{n+\frac{1}{2}+\sqrt{\frac{1}{4}+\tilde{\gamma}^2+H(H+1)}}\right]^2
\end{split}
\ea
for pseudospin symmetry limit. Their corresponding wave functions are
\ba
\begin{split}
F_{n,-1}(r)=N_{n}&e^{-r \vartheta}(1- e^{-2\delta r})^{(1+\zeta)/2} P_{n}^{ \left(\vartheta/\delta,\zeta \right)} (1 - 2e^{-2\delta r}),\\
G_{n,1}(r) = \tilde N_{n}& e^{-r \tilde \vartheta} (1-e^{-2\delta r})^{(1+\tilde{\zeta})/2} P_{n}^{\left( \tilde \vartheta/\delta,\tilde{\zeta} \right)}(1-2e^{-2\delta r}),
\end{split}
\label{FnkGnkSwave}
\ea
where we have introduce the following relations
\ba
\begin{split}
	\vartheta=& \sqrt{M^2-E_{n,-1}^2-C_S(M-E_{n,-1})}, \qquad	\zeta = \sqrt{1+4\gamma^2+4H(H-1)}, \\
	\tilde{\vartheta} =& \sqrt{M^2-E_{n,1}^2 + C_{PS}(M + E_{n,1})}, \qquad
	\tilde{\zeta} = \sqrt{1+4\tilde{\gamma}^2+4H(H+1)}.
\end{split}
\label{FnkGnkSwave+}
\ea

  \item[\textit{2. Dirac-Hulth\'en problem}:] For $V'_{0}=B=0$, our potential turns to the Hulth\'en potential, and the energy eigenvalue for the spin symmetry becomes
\ba
M^2-E_{n \kappa}^{2} - C_S(M - E_{n \kappa})= 4\delta^2\left[\frac{Z e^2(M+E_{n\kappa}-C_S)}{4\delta (n +\kappa+H+1)}-\frac{(n +\kappa+H+1)}{2}\right]^2,
\ea
while for the pseudospin symmetry
\ba
M^2-E_{n \kappa}^{2} + C_{PS}(M + E_{n \kappa})= \delta^2\left[\frac{-Z e^2(M-E_{n\kappa}+C_{PS})}{2\delta (n +\kappa+H)}-(n +\kappa+H)\right]^2.
\ea
In the condition of vanishing tensor interaction ($H=0$), these results turn out to be the same as the expressions obtained in Eq.(35) and Eq.(47) of Ref.~\cite{Soylu}, and also the results in Ref.~\cite{Ikhdair11}. The corresponding wave functions for both symmetry cases can be expressed as
\ba
\begin{split}
F_{n \kappa}(r)&=N_{n \kappa}e^{- r\xi}(1- e^{-2\delta r})^{\eta_\kappa+1} P_{n}^{\left(\xi / \delta, 2 \eta_\kappa + 1 \right) }(1-2 e^{-2\delta r}),\\
G_{n \kappa}(r) &= \tilde N_{n \kappa} e^{- r\tilde{\xi}} (1- e^{-2\delta r})^{\eta_\kappa}  P_{n}^{\left( \tilde{\xi}/\delta,2\eta_\kappa-1\right)}(1-2 e^{-2\delta r}),
\end{split}
\label{FnkGnkHulten}
\ea
with
\ba
\begin{split}
	\xi =\sqrt{M^2-E_{n\kappa}^2-C_S(M-E_{n\kappa})}, \qquad \tilde{\xi} = \sqrt{M^2-E_{n\kappa}^2 + C_{PS}(M + E_{n\kappa})}.
\end{split}
\label{FnkGnkHulten+}
\ea

  \item[\textit{3. Dirac-Yukawa problem}:] We also notice that setting $V_{0}=B=0$ gives us the energy spectrum for the Yukawa potential
\ba
M^2-E_{n \kappa}^{2} - C_S(M - E_{n \kappa}) = 4\delta^2\left[\frac{A(M+E_{n\kappa}-C_S)}{4\delta (n +\kappa+H+1)}-\frac{(n +\kappa+H+1)}{2}\right]^2
\ea
for the spin symmetry, and
\ba
M^2-E_{n \kappa}^{2} + C_{PS}(M + E_{n \kappa})= \delta^2\left[\frac{-A(M-E_{n\kappa}+C_{PS})}{2\delta (n +\kappa+H)}-(n +\kappa+H)\right]^2
\ea
for the pseudospin symmetry case. As $H=0$, these equations are identical to Eq.(30) and Eq.(31) of Ref.~\cite{Aydogdu11} for the spin, and Eq.(25) and Eq.(43) of Ref.~\cite{Ikhdair12} for the pseudospin symmetry case. Their spinor wave functions have the same form as Eq.\eqref{FnkGnkHulten} for each cases.

\item[\textit{4. Dirac-Coulomb-like problem}:] Taking the limit $\delta\rightarrow 0$ in the Yukawa potential, we obtain the well-known Coulomb-like potential $V(r)=-A/r$. The energy spectrum for both symmetry cases respectively yield
\ba
\begin{split}
(E_{n \kappa}-M)(E_{n \kappa}+M-C_S)= -\frac{A^2 (-C_S+E_{n \kappa}+M)^2}{4 (n+\kappa+H+1)^2}
\Rightarrow E^{S}_{n \kappa}=\frac{A^2 (C_S-M)+4 M (n+\kappa+H+1)^2}{A^2+4 (n+\kappa+H+1)^2},
\end{split}
\ea
\ba
\begin{split}
(E_{n \kappa}+M) (C_{PS}-E_{n \kappa}+M)=\frac{A^2 (C_{PS}-E_{n \kappa}+M)^2}{4 (n+\kappa+H)^2}
\Rightarrow E^{PS}_{n \kappa}=\frac{A^2 (C_{PS}+M)-4 M (n+\kappa+H)^2}{A^2+4 (n+\kappa+H)^2}.
\end{split}
\ea
For $H=0$, these results are respectively identical to Eqs.(59) and (60) of Ref.\cite{AydogduMie} with the replacement of $B\rightarrow A$. They are also in agreement with the results in Eqs.(53) and (56) of Ref.~\cite{Ikhdair12}. Moreover, if $C_S=C_{PS}=0$, these results reduce to the Dirac–Coulomb problem as
\ba
\begin{split}
&E^S_{n \kappa}=M\left[\frac{4(n+\kappa+1)^2-A^2}{4 (n+\kappa+1)^2+A^2}\right], ~~E^{PS}_{n \kappa}=-M\left[\frac{4(n+\kappa)^2-A^2}{4 (n+\kappa)^2+A^2}\right]
\end{split}
\ea
We note that the same expressions can also be achieved by taking the limit $\delta\rightarrow 0$ of the Hulth\'en potential under the replacement of $A \leftrightarrow Z e^2$.

  \item[\textit{5. Dirac-inversely quadratic Yukawa problem}:] When the parameters $V_{0}$ and $V'_{0}$ is fixed to zero, then we find the energy spectrum for inversely quadratic Yukawa potential as follows:
\ba
\begin{split}
M^2&-E_{n \kappa}^{2} - C_S(M - E_{n \kappa}) =\left[\frac{\eta_\kappa(\eta_\kappa+1)+\frac{1}{2}+n(n+1)+(2n+1)\sqrt{\frac{1}{4}+\gamma^2+\eta_\kappa(\eta_\kappa+1)}}{n+\frac{1}{2}+\sqrt{\frac{1}{4}+\gamma^2+\eta_\kappa(\eta_\kappa+1)}} \delta\right]^2
\end{split}
\ea
for spin symmetry limit and
\ba
\begin{split}
M^2&-E_{n \kappa}^{2} + C_{PS}(M + E_{n \kappa}) =\left[\frac{\eta_\kappa(\eta_\kappa-1)+\frac{1}{2}+n(n+1)+(2n+1)\sqrt{\frac{1}{4}+\tilde{\gamma}^2+\eta_\kappa(\eta_\kappa-1)}}{n+\frac{1}{2}+\sqrt{\frac{1}{4}+\tilde{\gamma}^2+\eta_\kappa(\eta_\kappa-1)}} \delta\right]^2
\end{split}
\ea
for pseudospin symmetry limit.
  \item[\textit{6. Dirac-Kratzer–Fues problem}:] By limiting $\delta \to 0$, the inversely quadratic Yukawa potential can be approximated as
\ba
V_{CY}=\lim_{\delta \to 0}\left(-\frac{A e^{-\delta r}}{r}-\frac{B e^{-2\delta r}}{r^2}\right)\simeq -\frac{A }{r}-\frac{B }{r^2},
\ea
where $A=2r_e D_e$ and $B=-r_e^2 D_e$. This form is well-known as the Kratzer–Fues potential. The energy spectrum from both symmetries according with this potential are
\ba
\begin{split}
M^2&-E_{n \kappa}^{2} - C_S(M - E_{n \kappa})=\frac{A^2 (M-C_S+E_{n \kappa})^2}{\left(2 n+1+2\sqrt{B (C_S-E_{n \kappa}-M)+(\eta_\kappa+\frac{1}{2})^2}\right)^2}
\end{split}
\ea
and
\ba
\begin{split}
M^2&-E_{n \kappa}^{2} + C_{PS}(M + E_{n \kappa})=\frac{A^2 (C_{PS}-E_{n \kappa}+M)^2}{\left(2 n+1+2\sqrt{B (C_{PS}-E_{n \kappa}+M)+(\eta_\kappa-\frac{1}{2})^2}\right)^2}
\end{split}
\ea
for spin and pseudospin symmetry cases, respectively. These results are exactly the same as Eq.(38) and (30) of Ref.\cite{HamzaviMie} for $C=0$, $A \rightarrow B$ and $B \rightarrow -A$.

\item[\textit{7. Non-relativistic limit}:] By setting $C_S=H=0$ and replacing $E_{n \kappa}+M \rightarrow 2m$ and $E_{n \kappa}-M \rightarrow E_{n l}$ in Eqs.\eqref{a17}, \eqref{a19} and \eqref{EnSS}, we have the non-relativistic solutions of the Hulth\'en plus a class of Yukawa potential. The resulting energy eigenvalue is
 \ba
\begin{split}
E_{n l}=-\frac{1}{2 m }\left[ \frac{\delta \left(-\frac{m \left(A+Z e^2 \right)}{ \delta}+(2 n+1)\sqrt{l^2+l+\frac{1}{4}-2m B }+l (l+1)+ n (n+1)+\frac{1}{2}\right)}{\sqrt{l^2+l+\frac{1}{4}-2m B }+n+\frac{1}{2}}\right]^2.
\label{EnlNonRel1}
\end{split}
\ea
Furthermore, setting $B=0$ simplifies the above equation as
\ba
E_{n l}=-\frac{1}{2 m }\left[\frac{\delta  (l+n+1)^2-m  \left(A+Ze^2\right)}{ (l+n+1)}\right]^2,~(n,l=0,1,2,\ldots),
\label{EnlNonRel}
\ea
which is coincide with Eq.(67) of Ref. \cite{Ikhdair112} by setting $d_0=0$, $V_0 \rightarrow A+Ze^2$ and $\delta \rightarrow 2\delta$. The same result is obtained in Eq.(27) of Ref. \cite{Haouat08} if we replace $\delta \rightarrow 2\delta$ and $\alpha \rightarrow A+Ze^2$. Furthermore, this is also identical to Eq.(35) of Ref.\cite{Ikhdair12} if we set $A \rightarrow A+Ze^2$.  We note that, considering the s-wave case ($l = 0$), Eq.\eqref{EnlNonRel} provides the exact result from the familiar nonrelativistic limit. Finally, when we set $\delta\rightarrow 0$ and $A=0$ in Eq.~\eqref{EnlNonRel}, the energy spectrum for the nonrelativistic Coulombic field is obtained as
\ba
E_{n l}=-\frac{m  \left(Ze^2\right)^2}{ 2(l+n+1)^2}.
\label{EnlNonRelCol}
\ea
\end{description}

\section{Numerical Evaluations and Discussion}\label{nr}
In this section, we perform the numerical evaluations for our analytical results. We analyze the dependency of the energy spectrum to the potential parameters for several quantum numbers. We can use an arbitrary unit to express the eigenvalues since the natural units are used in this study. Considering this issue, we prefer to use fm$^{-1}$ unit for the involved parameters in our calculation to obtain more realistic descriptions.
\begin{table}[htb]
\caption{Bound state energy eigenvalues (in fm$^{-1}$) of the spin symmetry case for various values of $n$ and $l$ in the absence ($H=0$) and presence ($H=5$) of the tensor interaction.}\label{table:BSss}
\centering
\begin{tabular}{ccccccccc}
\hline
$l$ & $n,\kappa<0$ & $(l,j=l+1/2)$ & $E_{n,\kappa<0}(H=0)$ & $E_{n,\kappa<0}(H=5)$& $n,\kappa>0$ & $(l,j=l-1/2)$ & $E_{n,\kappa>0}(H=0)$&$E_{n,\kappa>0}(H=5)$\\
\hline
1& 0,-2 & 0$p_{3/2}$ & 0.24181258 &0.24725816& 0,1 &  0$p_{1/2}$& 0.24181258&0.26229015\\
2& 0,-3 & 0$d_{5/2}$ & 0.24408024 &0.24408024& 0,2 &  0$d_{3/2}$& 0.24408024&0.26915052\\
3& 0,-4 & 0$f_{7/2}$ & 0.24725817 &0.24181258& 0,3 &  0$f_{5/2}$& 0.24725817&0.27694672\\
4& 0,-5 & 0$g_{9/2}$ & 0.25134955 &0.24045289& 0,4 &  0$g_{7/2}$& 0.25134955&0.28568689\\
1& 1,-2 & 1$p_{3/2}$ & 0.24407876 &0.25134791& 1,1 &  1$p_{1/2}$& 0.24407876&0.26914833\\
2& 1,-3 & 1$d_{5/2}$ & 0.24725666 &0.24725666& 1,2 &  1$d_{3/2}$& 0.24725666&0.27694432\\
3& 1,-4 & 1$f_{7/2}$ & 0.25134791 &0.24407876& 1,3 &  1$f_{5/2}$& 0.25134791&0.28568428\\
4& 1,-5 & 1$g_{9/2}$ & 0.25635671 &0.24181038& 1,4 &  1$g_{7/2}$& 0.25635671&0.29537745\\
1& 2,-2 & 2$p_{3/2}$ & 0.24725314 &0.25635387& 2,1 &  2$p_{1/2}$& 0.24725314&0.27694119\\
2& 2,-3 & 2$d_{5/2}$ & 0.25134496 &0.25134496& 2,2 &  2$d_{3/2}$& 0.25134496&0.28568098\\
3& 2,-4 & 2$f_{7/2}$ & 0.25635387 &0.24725314& 2,3 &  2$f_{5/2}$& 0.25635387&0.29537396\\
4& 2,-5 & 2$g_{9/2}$ & 0.26228527 &0.24407133& 2,4 &  2$g_{7/2}$& 0.26228527&0.30603052\\
1& 3,-2 & 3$p_{3/2}$ & 0.25133807 &0.26228075& 3,1 &  3$p_{1/2}$& 0.25133807&0.28567666\\
2& 3,-3 & 3$d_{5/2}$ & 0.25634875 &0.25634875& 3,2 &  3$d_{3/2}$& 0.25634875&0.29536954\\
3& 3,-4 & 3$f_{7/2}$ & 0.26228075 &0.25133807& 3,3 &  3$f_{5/2}$& 0.26228075&0.30602597\\
4& 3,-5 & 3$g_{9/2}$ & 0.26914103 &0.24723546& 3,4 &  3$g_{7/2}$& 0.26914103&0.31765756\\
\hline
\end{tabular}
\end{table}
\begin{table}[ht]
\caption{Bound state energy eigenvalues (in fm$^{-1}$) of the pseudospin symmetry case for various values of $n$ and $\widetilde{l}$ in the absence ($H=0$) and presence ($H=5$) of the tensor interaction.}\label{table:BSps}
\centering
\begin{tabular}{ccccccccc}
\hline
$\widetilde{l}$ & $n,\kappa<0$ & $(l,j)$ & $E_{n,\kappa<0}(H=0)$& $E_{n,\kappa<0}(H=5)$& $n-1,\kappa>0$ & $(l+2,j+1)$ & $E_{n,\kappa>0}(H=0)$& $E_{n,\kappa>0}(H=5)$\\
\hline
1& 1,-1 & 1$s_{1/2}$ &-0.24665137 &-0.25853490& 0,2 &  0$d_{3/2}$&-0.24665137&-0.28786907\\
2& 1,-2 & 1$p_{3/2}$ &-0.25183976 &-0.25183976& 0,3 &  0$f_{5/2}$&-0.25183976&-0.30082457\\
3& 1,-3 & 1$d_{5/2}$ &-0.25853490 &-0.24665137& 0,4 &  0$g_{7/2}$&-0.25853490&-0.31543002\\
4& 1,-4 & 1$f_{7/2}$ &-0.26675519 &-0.24295721& 0,5 &  0$h_{9/2}$&-0.26675519&-0.33173176\\
1& 2,-1 & 2$s_{1/2}$ &-0.25184913 &-0.26676283& 1,2 &  1$d_{3/2}$&-0.25184913&-0.30083316\\
2& 2,-2 & 2$p_{3/2}$ &-0.25854279 &-0.25854279& 1,3 &  1$f_{5/2}$&-0.25854279&-0.31543915\\
3& 2,-3 & 2$d_{5/2}$ &-0.26676283 &-0.25184913& 1,4 &  1$g_{7/2}$&-0.26676283&-0.33174151\\
4& 2,-4 & 2$f_{7/2}$ &-0.27653162 &-0.24667097& 1,5 &  1$h_{9/2}$&-0.27653162&-0.34979406\\
1& 3,-1 & 3$s_{1/2}$ &-0.25856120 &-0.27654386& 2,2 &  2$d_{3/2}$&-0.25856120&-0.31545110\\
2& 3,-2 & 3$p_{3/2}$ &-0.26677657 &-0.26677657& 2,3 &  2$f_{5/2}$&-0.26677657&-0.33175386\\
3& 3,-3 & 3$d_{5/2}$ &-0.27654386 &-0.25856119& 2,4 &  2$g_{7/2}$&-0.27654386&-0.34980694\\
4& 3,-4 & 3$f_{7/2}$ &-0.28788894 &-0.25189558& 2,5 &  2$h_{9/2}$&-0.28788894&-0.36967266\\
1& 4,-1 & 4$s_{1/2}$ &-0.25856120 &-0.28790739& 3,2 &  3$d_{3/2}$&-0.25856120&-0.33177001\\
2& 4,-2 & 4$p_{3/2}$ &-0.26677657 &-0.27656587& 3,3 &  3$f_{5/2}$&-0.26677657&-0.34982325\\
3& 4,-3 & 4$d_{5/2}$ &-0.27654386 &-0.26680859& 3,4 &  3$g_{7/2}$&-0.27654386&-0.36968936\\
4& 4,-4 & 4$f_{7/2}$ &-0.28788894 &-0.25865185& 3,5 &  3$h_{9/2}$&-0.28788894&-0.39144042\\
\hline
\end{tabular}
\end{table}

In Table \ref{table:BSss} and  \ref{table:BSps}, we present several energy levels $E_{n,\kappa}$ for the case of spin symmetry and pseudospin symmetry. We perform this calculation by using Eq.\eqref{EnSS} for the spin symmetry case and Eq.\eqref{EnkPS} for the pseudospin symmetry case. The outcomes include the absence as well as presence of the tensor coupling. In the calculation, we set $C_S = 5$ fm$^{-1}$, $C_{PS}=-5$ fm$^{-1}$, $A=B=1$ fm$^{-1}$, $V_0=2$ fm$^{-1}$, and $\delta=0.05$ fm$^{-1}$ for convenient. These parameters can vary according to the considered bound states and here they solely represent the widely used benchmarks for numerical purposes. As for the nucleon mass, the corresponding value is $M=939$ MeV $\approx 4.76$ fm$^{-1}$. We have chosen these values to meet the appropriate range of nuclear studies, particularly related to the single-nucleon states. From both tables, we notice that $E_{n,\kappa}$ increases as the increment of $|\kappa|$ on both symmetry considerations for a given $n$. The absence of tensor interaction ($H=0$) on the spin symmetry case evoke degeneracy in some Dirac spin-doublet eigenstate partners: ($np_{3/2}, np_{1/2}$), ($nd_{5/2}, nd_{3/2}$), ($nf_{7/2}, nf_{5/2}$), and ($ng_{9/2}, ng_{7/2}$), etc. Each of these two spin-doublet pairs has the same $n,l$. Under the same case, degeneracy also occur on the pseudospin symmetry in some pseudospin-doublet partners: ($ns_{1/2}$,$(n-1)d_{3/2}$), ($np_{3/2}$,$(n-1)f_{5/2}$), ($nd_{5/2}$,$(n-1)g_{7/2}$), and ($nf_{7/2}$,$(n-1)h_{9/2}$), etc. Again, each of these two states has the same $\tilde{n}$ and $\tilde{l}$. However, as the tensor interaction appears, all these degeneracies on both symmetry considerations vanish.

\begin{figure*}[hbt]
    \begin{center}
\includegraphics[scale=0.38]{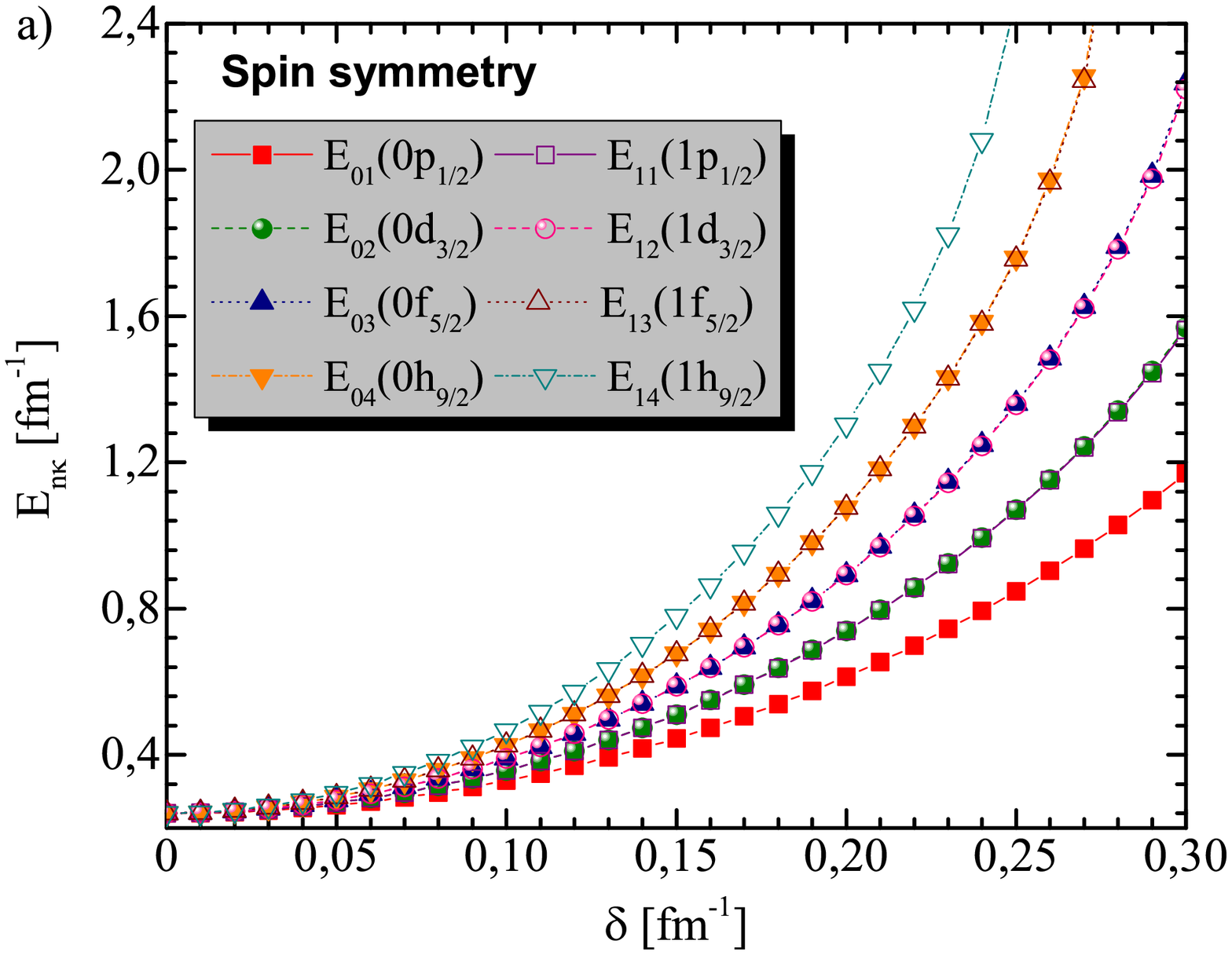}
\includegraphics[scale=0.38]{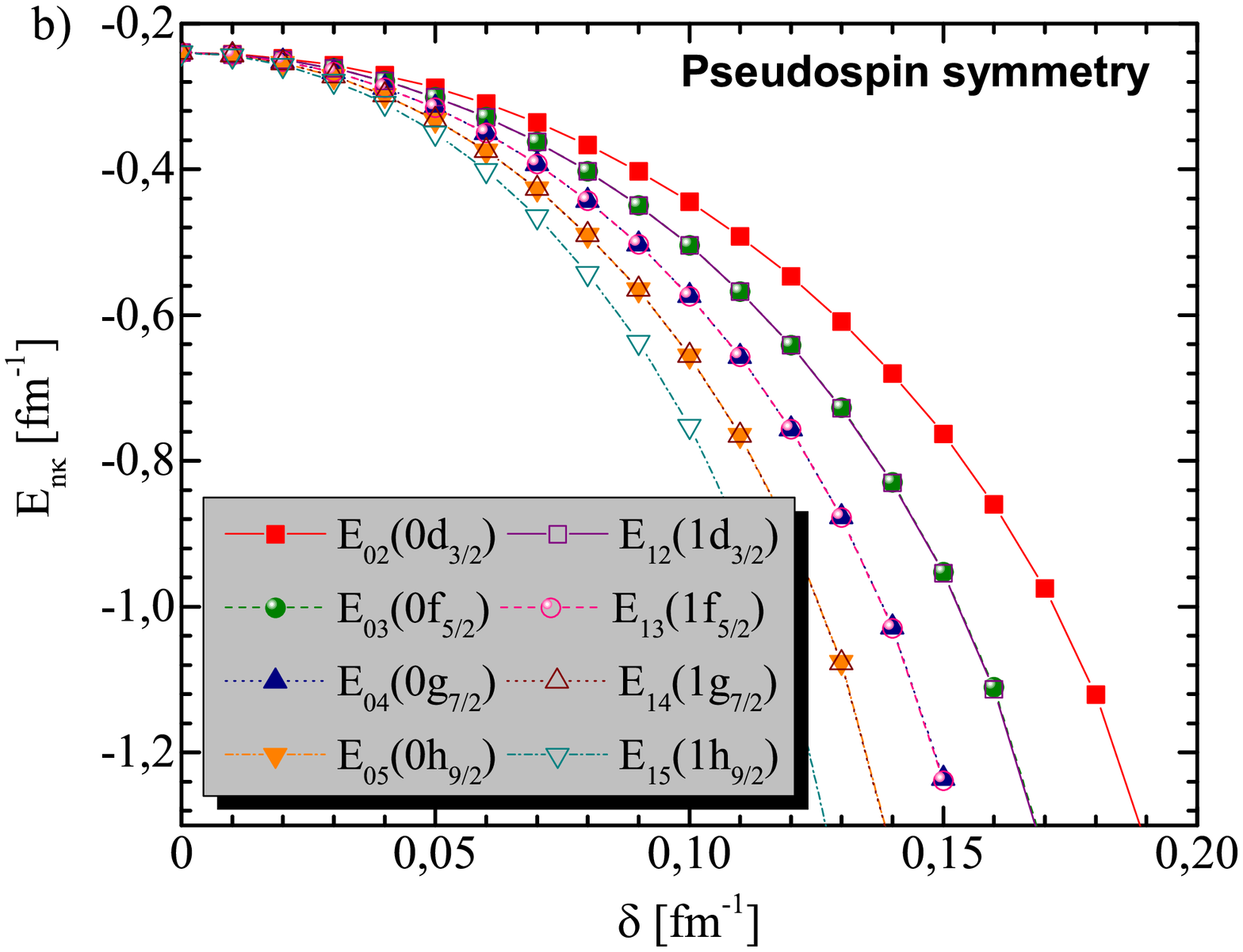}
     \end{center}
          \vspace{-4mm}
\caption{ The variation of $E_{n=0,1,\kappa=1,2,3,4}$ with respect to screening parameter $\delta$ for (a) spin and (b) pseudospin symmetry case with $H=5$, $C_{S}=5$ fm$^{-1}$, $C_{PS}=-5$ fm$^{-1}$, $M=4.76$ fm$^{-1}$, $A=B=1$ fm$^{-1}$ and $V_0=2$ fm$^{-1}$.}
\label{fig:Edelta}
\end{figure*}

We present the dependence of $E_{n\kappa}$ on $\delta$ for different $n$ and $\kappa$ by setting the other parameters with the previous benchmarks in Fig.\ref{fig:Edelta} for the (a) spin and (b) pseudospin symmetry case. The behavior of $E_{n\kappa}$ is demonstrated by varying $\delta$ from $0$ to $0.30$ fm$^{-1}$ with $0.01$ fm$^{-1}$ step. Note that increasing the value of $\delta$ implies the less attractive interactions. As $\delta$ rises for a short-range potential, the bound state energy eigenvalues increase for the spin symmetry and decrease for the pseudospin symmetry case. The increasing trend indicates that we have tightly bounded states, while the decreasing behavior means otherwise.
\begin{figure}[hbt]
    \begin{center}
\includegraphics[scale=0.39]{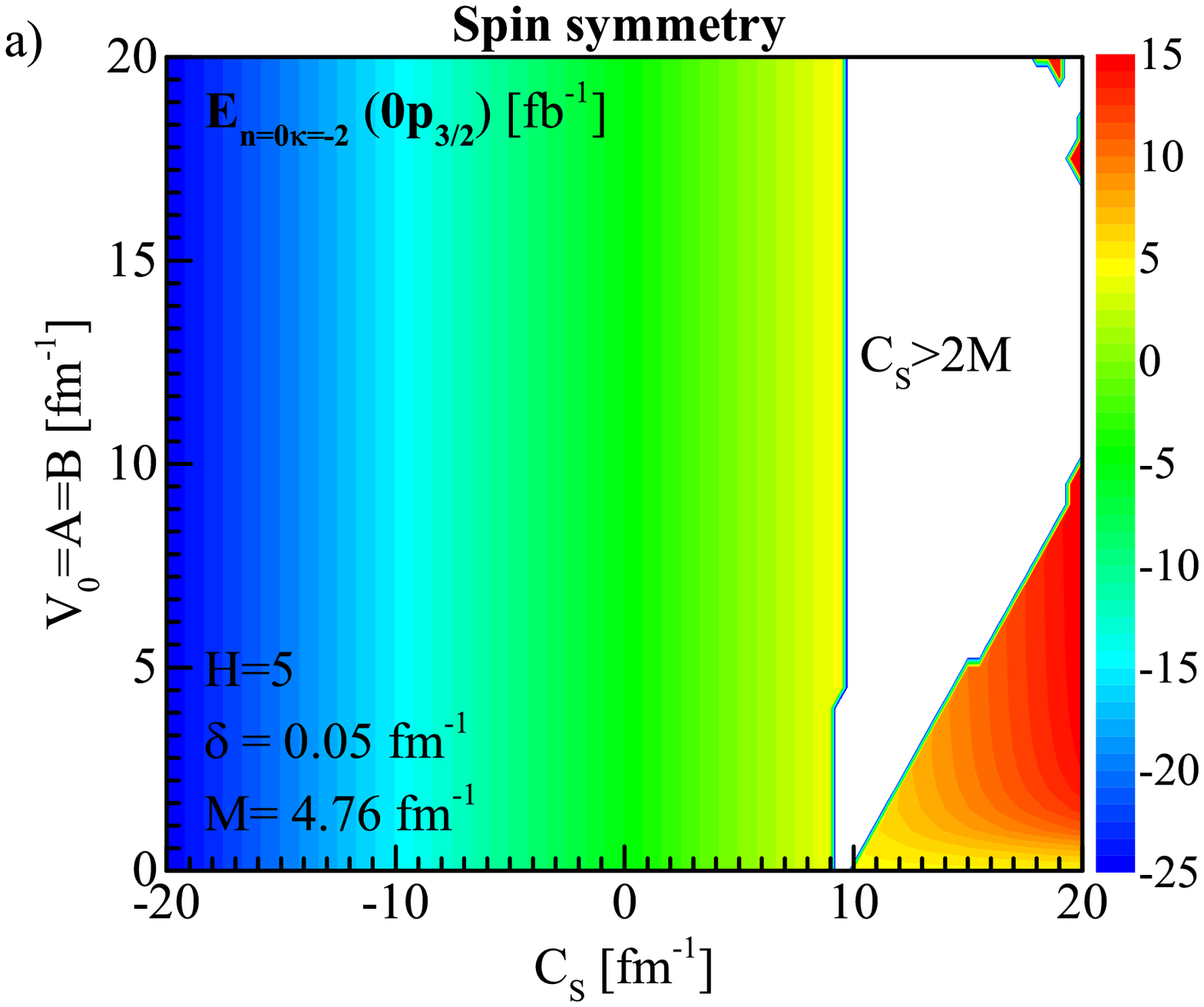}
\includegraphics[scale=0.39]{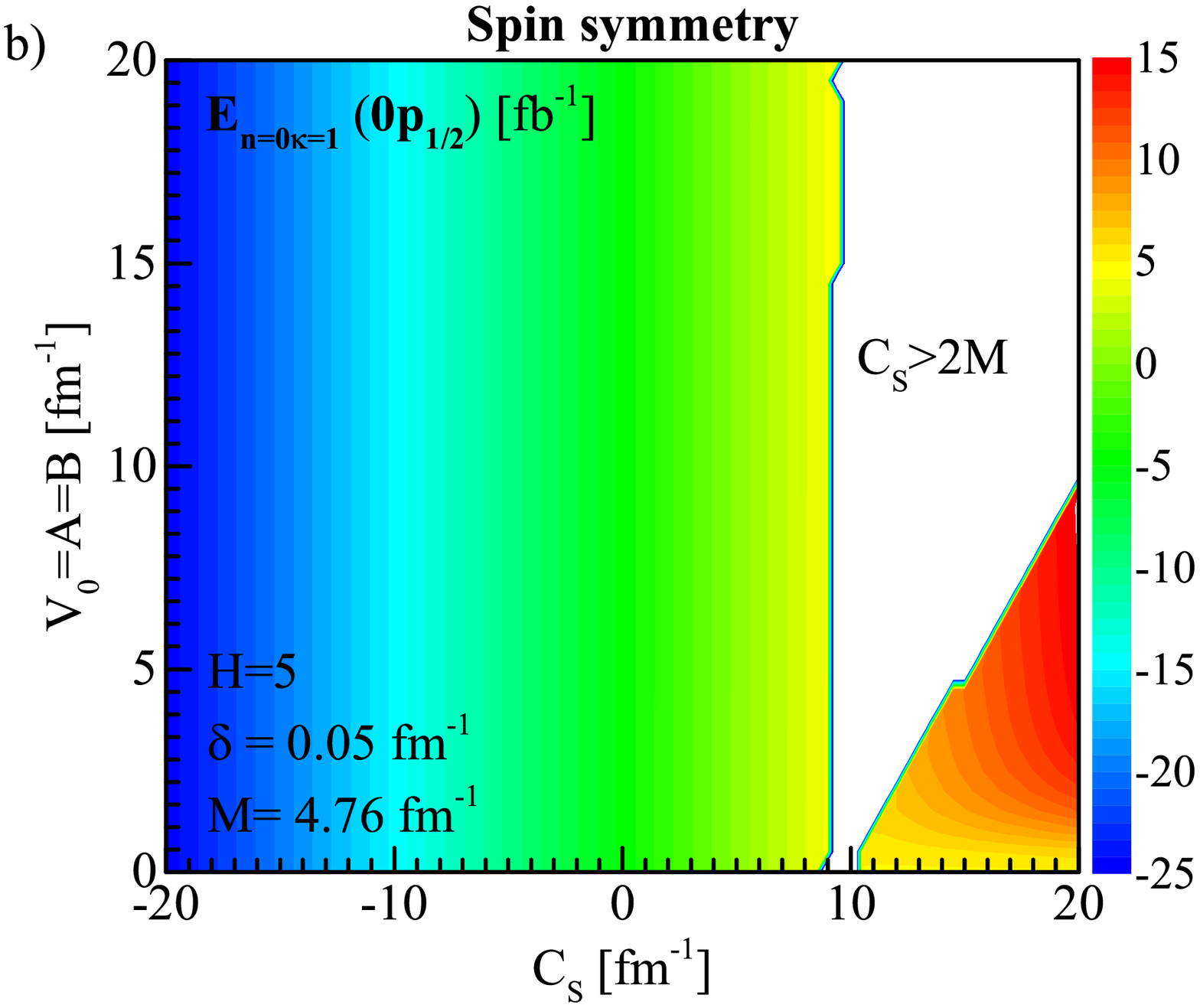}
\includegraphics[scale=0.39]{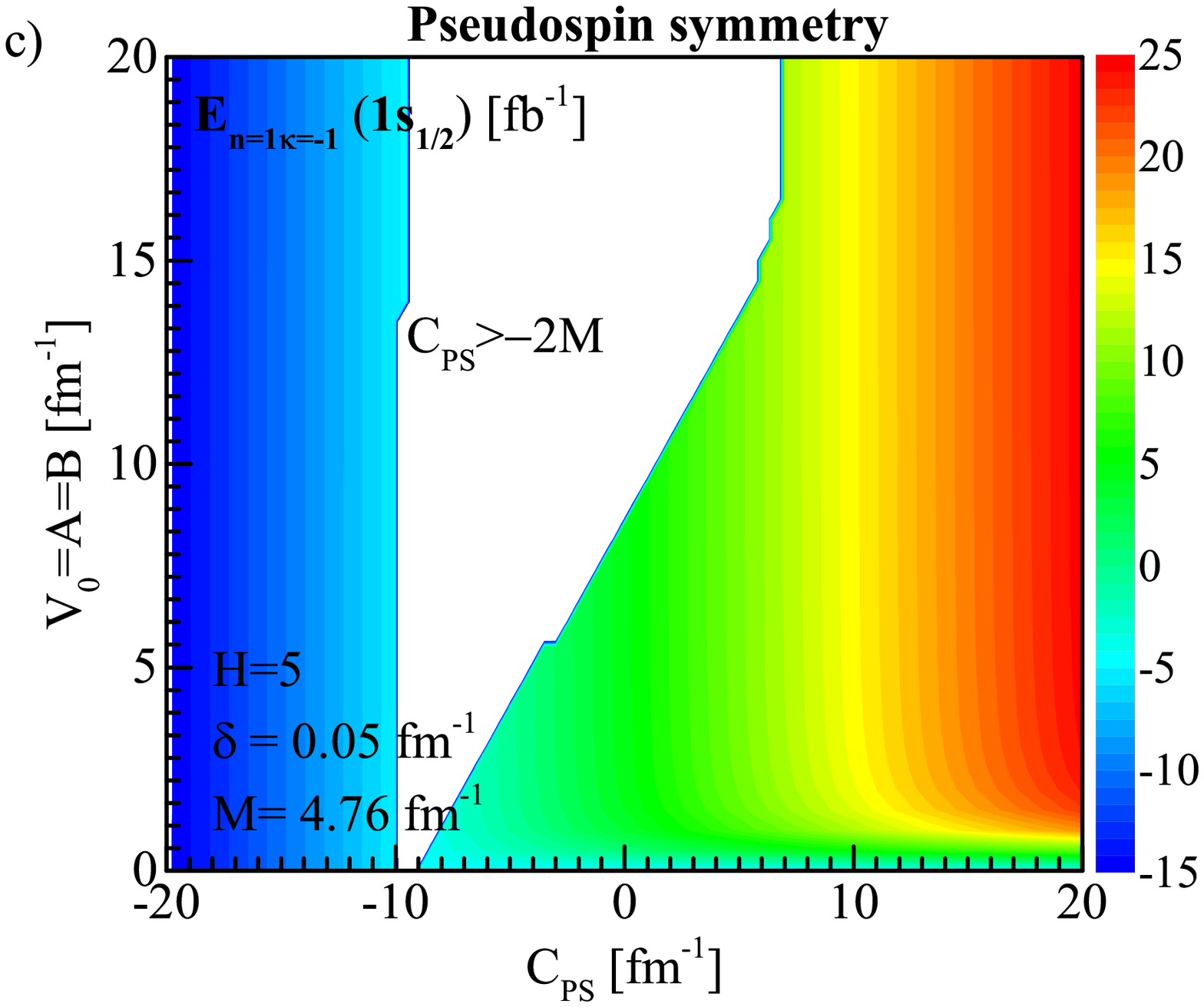}
\includegraphics[scale=0.39]{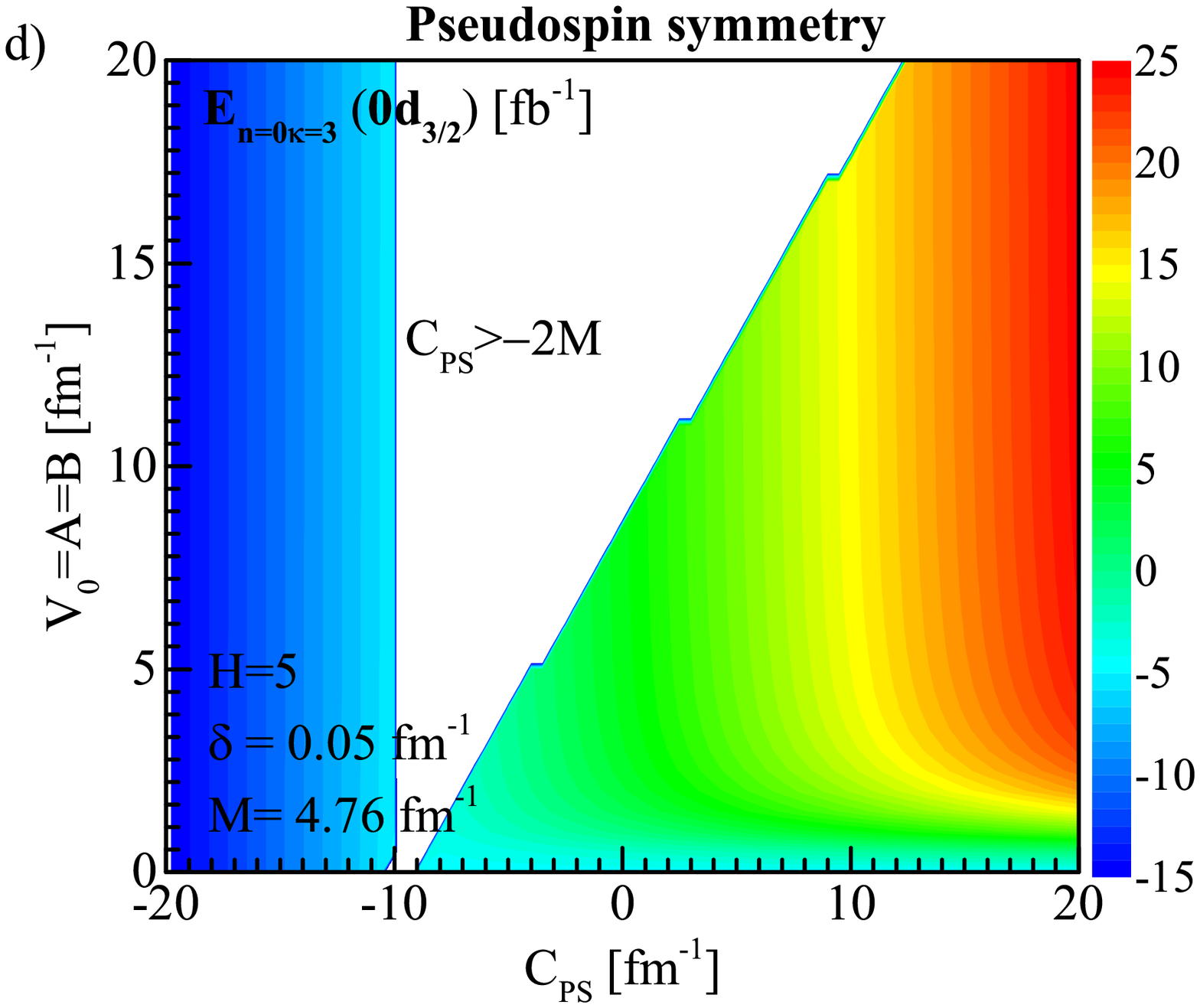}
     \end{center}
     \vspace{-4mm}
\caption{The energy eigenvalues of the spin symmetry states (a) $0p_{3/2}$ and  (b) $0p_{1/2}$, and the pseudospin symmetry states (c) $0s_{1/2}$ and (d) $0d_{3/2}$ in the plane of $(V_0,C_{S,PS})$. The color heat map corresponds to the energy eigenvalue (in fm$^{-1}$) of the scanned region.}
\label{fig:EVC}
\end{figure}
Particularly, we can see that the binding energies are stepped away from each other as $\delta$ increases in both symmetries. We can see that the lines of the following pair of states overlap: ($0d_{3/2},1p_{1/2}$), ($0f_{5/2},1d_{3/2}$) and ($0g_{7/2},1f_{5/2}$) for the spin symmetry case, and ($0f_{5/2},1d_{3/2}$), ($0g_{7/2},1f_{5/2}$), and ($0h_{9/2},1g_{7/2}$) for the pseudospin symmetry case.

In Fig.\ref{fig:EVC}, we illustrate the parameter space of the energy eigenvalues for the spin symmetry states  $0p_{3/2}$ and $0p_{1/2}$, and the pseudospin symmetry states $0s_{1/2}$ and $0d_{3/2}$. The energy eigenvalues are scanned in $(V_0,C_{S})$ and $(V_0,C_{PS})$ plane for each cases, respectively. For this objective, we set $V_0=A=B$, $H=5$, $\delta=0.05$ fm$^{-1}$, and $M=4.76$ fm$^{-1}$. The scan parameters are varied from $0.0$ to $20.0$ fm$^{-1}$ for $V_0$, and from $-20.0$ to $20.0$ fm$^{-1}$ for $C_{S,PS}$ with $0.5$ fm$^{-1}$ step. The white region represents non-real energy eigenvalue. That is, there are no bound states occur in this domain. It is clear that the energy spectrums depend entirely on the choice of $C_{S}$ and $C_{PS}$. The positive bound state of the spin symmetry case are obtained at the regions of $5\leq C_{S}< 10$ with $M\geq E_{n \kappa}$ and $E_{n \kappa}+M\geq C_S$, as well as of $10\leq C_{S}\leq 20$ with $M< E_{n \kappa}$ and $E_{n \kappa}+M\leq C_S$. Meanwhile, the negative bound state energy eigenvalues for the pseudospin symmetry case are reached at the regions of $-10<C_{PS}\leq-5$ with $M>-E_{n \kappa}$ and $E_{n \kappa}<C_{PS}+M$, as well as of $-20<C_{PS}\leq-10$ with $M<-E_{n \kappa}$ and $E_{n \kappa}<C_{PS}+M$. In addition, these results are valid for other quantum states from the same distributions as the ones we have discussed.

\begin{figure*}[!hbt]
    \begin{center}
\includegraphics[scale=0.38]{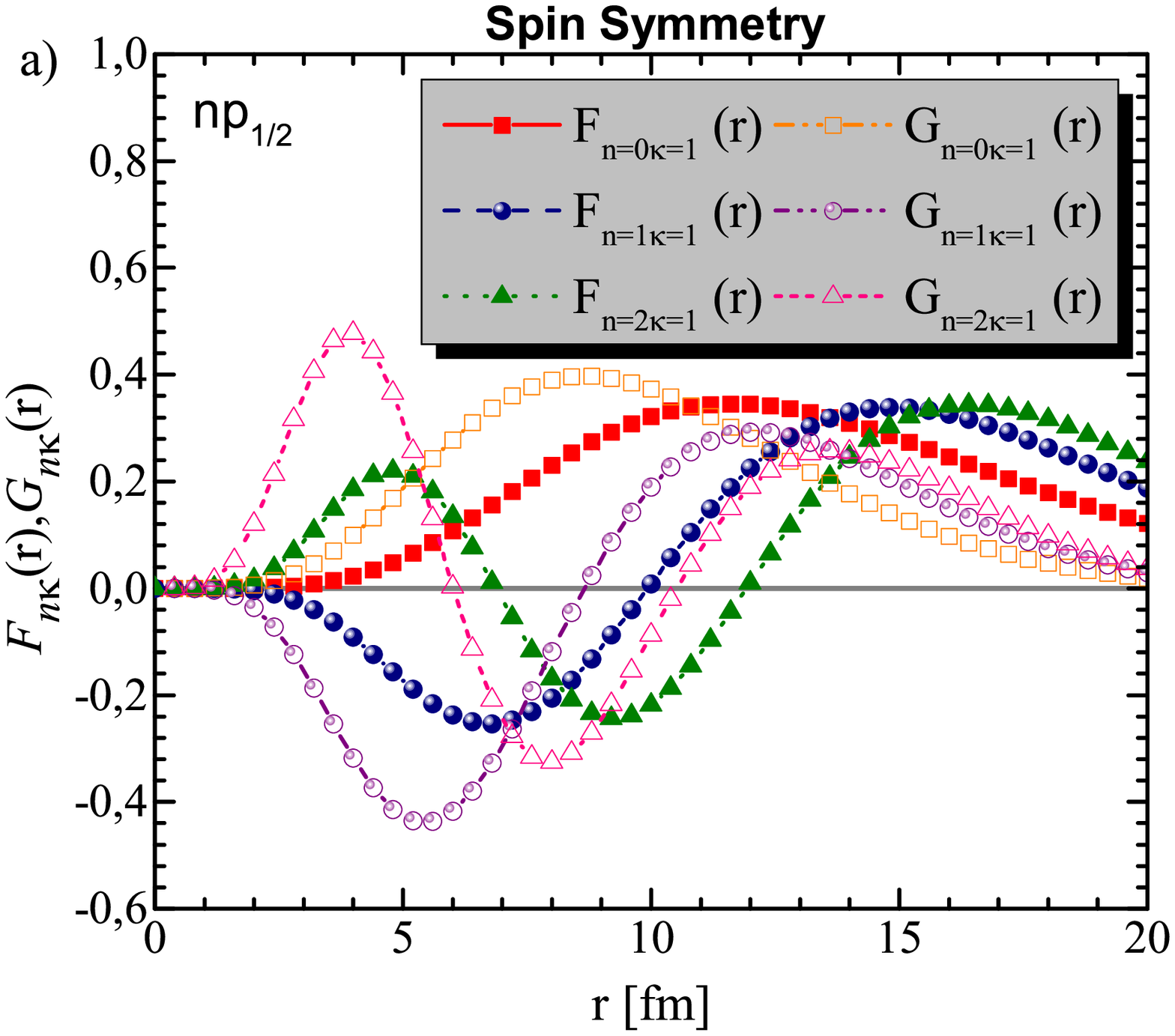}
\includegraphics[scale=0.38]{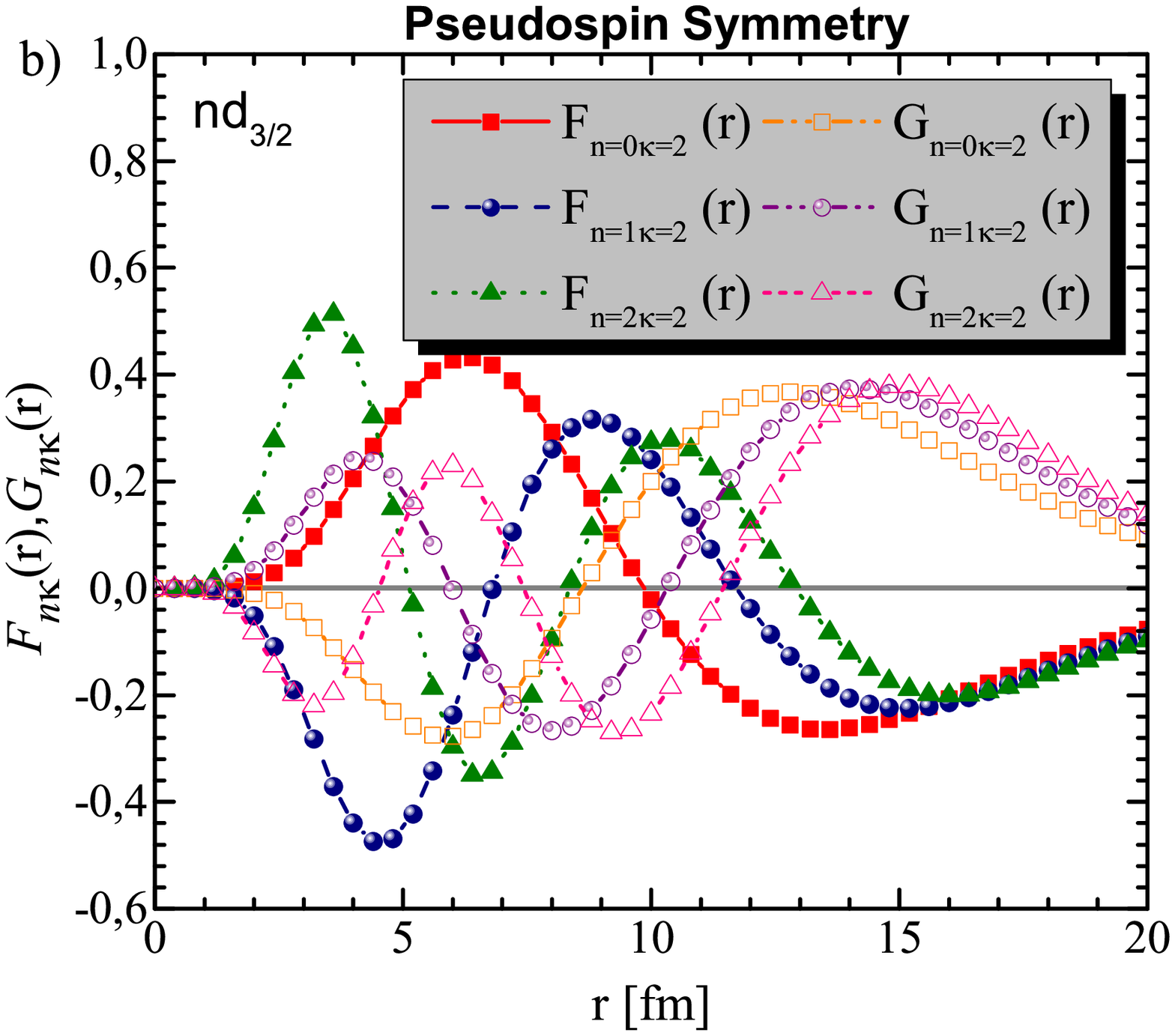}
     \end{center}
     \vspace{-4mm}
\caption{Variation of the normalized upper and lower  components of $F_{n,\kappa}(r)$ and $G_{n,\kappa}(r)$ with respect to $r$ for (a) spin and (b) pseudospin symmetry consideration with $H=5$.}
\label{fig:FnkGnk}
\end{figure*}
Finally, we present the lower and upper spinor wave functions of $np_{1/2}$ and $nd_{3/2}$ states as a function of $r$ with $n=0,1,2$ in Fig. \ref{fig:FnkGnk} for (a) spin and (b) pseudospin symmetry case. We have implemented Eqs.\eqref{a50}, \eqref{GnkSS2}, \eqref{a87} and \eqref{FnkPS2} for these purposes. We use the same parameter values as in the Tables \ref{table:BSss} and \ref{table:BSps}. We can see that, the lower and upper component of the spin symmetry case have $n$ nodes, while for the pseudospin $n+1$ nodes. The $r$ dependence of the potential strengths (i.e. $V_0$, $A$, $B$) keeps the number of radial nodes stays the same, yet still influencing the wavelength and magnitude of the appropriate solutions.

\section{Summary and Conclusions}\label{cr}
In this work, we have examined the bound state solutions of the Dirac equation with a new suggested combined potential, the Hulth\'en plus a class of Yukawa potential, as well as including a Coulomb-like tensor coupling under the conditions of the spin and pseudospin symmetry. For this subject, we have implemented the NU and SUSYQM methods. The tensor coupling preserves the form of the combined potential, however produces a new spin-orbit centrifugal term $\eta(\eta\pm1) r^{-2}$ where $\eta$ denotes a new spin-orbit quantum number. It provides the possibility of establishing a different form of spin-orbit coupling terms that may evoke some physical interpretation.

For an arbitrary spin-orbit coupling quantum number $\kappa$, we have obtained analytical expressions for the energy eigenvalues and its associated upper- and lower-spinor wave functions in the spin as well as pseudospin symmetry cases. Results from the two methods are entirely the same. Both are systematic and practical for solving the considered symmetries and considered as two of the most reliable methods in this subject in many cases. The wave functions are expressed in terms of the hypergeometric functions, together with their normalization constants. Although the energy spectrums overlap with each other, the obtained wave functions from the SUSYQM are more compact than those from the NU method. Hence, the validity of the SUSYQM method and its general principles has been confirmed.

Furthermore, we have shown that our obtained results can be reduced into a few special cases (s-wave case, Dirac-Hulth\'en problem, Dirac-Yukawa problem, Dirac-Coulomb-like problem, Dirac-inversely quadratic Yukawa problem, Dirac-Kratzer–Fues problem) and compared them with the literature. They are in full consistency with the previous findings. Additionally, we have also considered the nonrelativistic limit of the energy spectrum for the proposed potential by making some replacements on the spin symmetry solution.

We have numerically investigated the dependence of the energy spectra dependence on the screening parameter $\delta$, potential strength, as well as parameter $C_S$ and $C_{PS}$. We found that both spin and pseudospin bound state energies are sensitive with $\delta$, as well as with $C_S$ and $C_{PS}$ for a given quantum number $\kappa$ and $n$. In the absence of the tensor coupling, the Dirac spin and pseudospin-doublet eigenstate partners evoke degeneracy for some states. However, the degeneracies are completely eliminated if the tensor interaction involved. The allowed bound state regions for both symmetries in the parameter space of the potential strength $V_0$ with respect to $C_S$, and $C_{PS}$ are also presented. Finally, the normalized wave function components from both symmetries, influenced with tensor interaction, are shown as a function of $r$.

In conclusion, a new suggested combined potential, Hulth\'en plus a class of Yukawa potential including a Coulomb-like tensor interaction, have been analytically solved. Our obtained results deserve particular attention to find relevancy in more applicative branches of physics, especially in the area of hadronic and nuclear physics.

\appendix
\section{Nikiforov-Uvarov Method}\label{NUformalism}
We briefly introduce the NU method \cite{Nikiforov}, a useful way to solve a hypergeometric-type second-order differential equation by transforming it into the following form
\begin{equation}
\frac{d^2\chi(s)}{ds^2} + \frac{\tilde{\tau}(s)}{\sigma(s)} \frac{d\chi(s)}{ds} + \frac{\tilde{\sigma}(s)}{\sigma^2(s)} \chi(s) =0.
\label{NU}
\end{equation}
All coefficients here are polynomials, in which $\sigma(s)$ and $\tilde{\sigma}(s)$ have a maximum second-order while $\tilde{\tau}(s)$ has a first-order kind. To get a particular solution for the above equation, the function $\chi(s)$ can be decomposed as
\begin{equation}
\chi(s) = y(s) \phi(s),
\label{NUfact}
\end{equation}
and then by substituting this into Eq.~\eqref{NU}, we find  a hypergeometric-type equation as follows
\begin{equation}
\sigma(s) \frac{d^2 y(s)}{ds^2} + \tau(s) \frac{dy(s)}{ds} + \lambda y (s) = 0.
\label{NUhyper}
\end{equation}

The function $\phi (s)$ need to satisfy
\begin{equation}
\frac{1}{\phi(s)}\frac{d\phi(s)}{ds} = \frac{\pi(s)}{\sigma(s)},
\label{NUphi}
\end{equation}
with
\begin{equation}
\pi(s) = \frac{\sigma'(s)-\tilde{\tau}(s)}{2} \pm \sqrt{\left[\frac{\sigma'(s)-\tilde{\tau}(s)}{2}\right]^2 -\tilde{\sigma}(s) + k\sigma(s)},
\label{NUphii}
\end{equation}
where primes denote the derivative according to $s$ and it can be first-order at most. The term within the square root is rearranged as zero discriminant of a second-order polynomial. Therefore, an expression for $k$ is found after solving such equation by means of the NU method.

Consequently, the equation reduces to a hypergeometric type equation, where one of its solutions is $y(s)$. Hence the polynomial expression $\bar\sigma(s)=\tilde\sigma(s)+\pi^2(s)+\pi(s)[\tilde\tau(s)-\sigma^{'}(s)]+\pi^{'}(s)\sigma(s)$
can be divided by a factor of $\sigma(s)$, such that $\bar\sigma/\sigma(s)=\lambda
$. Here, we use the following relation
\ba \lambda=k+\frac{d\pi(s)}{ds}, \label{NUgeneigen}
\ea
\begin{equation}
\tau(s) = \tilde{\tau}(s) + 2\pi(s),
\end{equation}
with $\tau(s)$ has a negative derivative. For an integer $n\geq0$, a unique $n$-degree polynomial solution is
obtained for the hypergeometric type equation if
\ba
\lambda \equiv \lambda_{n}=-n\frac{d\tau}{ds}-\frac{n(n-1)}{2}\frac{d^2\sigma}{ds^2}, \qquad n=0,1,2...
 \label{NUeigen}
\ea
On the other hand, the polynomial $y(s)$ satisfies the following Rodrigues equation
\begin{equation}
y_n(s) = \frac{C_n}{\rho(s)}\frac{d^n}{ds^n}\Big[\rho(s)\sigma^n(s)\Big].
\label{NUrod}
\end{equation}
The parameter $C_n$ denotes the normalization constant, while $\rho(s)$ stands for weighting function that obeys
\begin{equation}
\frac{d\left[\sigma(s) \rho(s)\right]}{ds} = \tau(s)\rho(s),
\label{NUweight}
\end{equation}
which is commonly known as the Pearson differential equation.

\end{document}